\documentclass[%
preprint,
 amsmath,amssymb,
 aps,
]{revtex4-2}

\usepackage{graphicx}
\usepackage{dcolumn}
\usepackage{bm}
\usepackage{xcolor}

\begin{document}

\preprint{APS/123-QED}

\title{Spreading and engulfment of a viscoelastic film onto a newtonian droplet}

\author{Chunheng Zhao}
 \affiliation{Sorbonne Universit\'{e} and CNRS, Institut Jean Le Rond d'Alembert UMR 7190, F-75005 Paris, France}
\author{Taehun Lee}
\affiliation{Department of Mechanical Engineering, City College of New York, New York, NY 10031, USA.
}%

\author{Andreas Carlson}%
 \email{acarlson@math.uio.no}
\affiliation{%
 Department of Mathematics, Mechanics Division, University of Oslo, Oslo 0316, Norway
}%

\date{\today}

\begin{abstract}
We use the conservative phase-field lattice Boltzmann method to investigate the dynamics when a Newtonian droplet comes in contact with an immiscible viscoelastic liquid film. The dynamics of the three liquid phases are explored through numerical simulations, with a focus on illustrating the contact line dynamics and the viscoelastic effects described by the Oldroyd-B model. The droplet dynamics are contrasted with the case of a Newtonian fluid film. The simulations demonstrate that when the film is viscoelastic, the droplet dynamics become insensitive to the film thickness when the polymer viscosity and relaxation time are large. A viscoelastic ridge forms at the moving contact line, which evolves with a power-law dependence on time. By rescaling the interface profile of the ridge using its height and width, it appears to collapse onto a similar shape. Our findings reveal a strong correlation between the viscoelastic stress and the interface shape near the contact line.
\end{abstract}

\maketitle

\section{Introduction}
Viscoelastic fluids can emerge from the mix of viscous solvents and elastic polymers. These Non-Newtonian fluids are found in industrial applications and biological phenomena, such as in 3D printing~\cite{duty2018makes}, fire safety~\cite{jaffe2015safer,wei2015megasupramolecules}, mucus~\cite{johansson2013gastrointestinal,lai2009micro,bansil2018biology}, tissue~\cite{forgacs1998viscoelastic,woodard1986composition}, cell rheology~\cite{janmey1991viscoelastic}, and natural convection~\cite{perez2018applications}, to name but a few examples. These polymers provide the fluid with an elastic property, causing it to resist deformations. As the viscoelastic fluids undergo deformation, the elastic polymers within them stretch thus creating a force. In the context of multiphase fluid flows, viscoelastic fluids introduce effects shown to determine the dynamics in processes like droplet coalescence~\cite{varma2022elasticity,dekker2022elasticity}, Plateau-Rayleigh instability~\cite{snoeijer2020relationship,wang2023viscoelastic,turkoz2018axisymmetric,li2023dynamics}, and droplet wetting phenomena~\cite{bouillant2022rapid,henkel2021gradient,greve2023stick,yada2023rapid}. 

When we place a droplet onto a viscous film of another immiscible liquid, the surface tension between droplet-air $\sigma_{wa}$, droplet-film $\sigma_{wv}$, and film-air $\sigma_{va}$ will induce contact line motion. Post-contact between the drop and the film, the interfacial dynamics can be decomposed into three parts ~\cite{cuttle2021engulfment,zhao2023engulfment}. First, the droplet deforms during a spreading-like motion as it extends along the film/wall. Second, the film progressively climbs up the droplet and may entirely cover its interface. Third, the droplet is pushed into the film by capillarity. The spreading factor $S_v=\sigma_{wa}-\sigma_{wv}-\sigma_{va}$ determines the final interfacial shape~\cite{pannacci2008equilibrium,carlson2013short,zhao2023engulfment}. Theoretically, when $S_v>1$, one would expect the droplet to be fully engulfed by the liquid film. In the case of Newtonian liquids, these dynamics have been described by both experiments and numerical simulations and the process of engulfment is affected by the height of the liquid film/pool~\cite{cuttle2021engulfment,zhao2023engulfment}. A key dimensionless number to describe these dynamics is the Ohnesorge number $Oh=\eta_n/\sqrt{\rho_n \sigma_{wn}D}$, representing the ratio between viscosity and inertia-capillarity, where $\rho_n$ is the density, $\eta_n$ is the viscosity where the subscript $n$ denotes the Newtonian liquid and $D$ the droplet diameter. 

Imagine now the same scenario, but the Newtonian fluid in the film is replaced by a viscoelastic fluid. The polymers in the viscoelastic fluid generate an elastic stress, which will affect the dynamics. From a mathematical aspect, the difference between Newtonian fluids and viscoelastic fluids enters into the stress tensor in the Navier-Stokes momentum equations~\cite{joseph2013fluid}. In addition to the Newtonian viscous contribution there is an additional nonlinear polymeric contribution in the viscoelastic fluid~\cite{snoeijer2020relationship}. While the general effects of viscoelasticity are well-documented, a detailed description of the spreading and engulfment dynamics specific to viscoelastic fluids is missing, which we address in this article by deploying numerical simulations of the three-phase liquid flow with viscoelastic effects.

The description of viscoelastic fluid flow is often based on models such as the upper-convected Maxwell  model~\cite{bird1977dynamics}, the Oldroyd B model, and the FENE P model~\cite{alves2021numerical}. All of these are used to describe the viscoelastic effect and its characteristics. Within these models, there are two important physical parameters, i.e., the polymer viscosity $\eta_p$ and the polymer relaxation time $\lambda_p$, characterizing the viscoelastic properties. The polymer viscosity, $\eta_p$, introduces a time-dependent strain rate, dissipating parts of energy within the system. Whereas the polymer relaxation time, $\lambda_p$, is the time scale required for the system to return to its equilibrium state or, in simpler terms, how long it takes for stress to relax. If letting these two parameters go towards infinity while maintaining a constant ratio, known as the shear modulus $G$, these models have been proposed to model viscoelastic solid systems~\cite{snoeijer2020relationship}. 
Many numerical schemes have been used to tackle the complex flow of viscoelastic fluids~\cite{alves2021numerical}. Initially, these governing equations were solved by a continuous differential equation for the polymeric stress $\boldsymbol{\Pi_p}$~\cite{thien1977new, giesekus1982simple}. Another computational approach focuses instead on the conformation tensor, which is a geometric tensor that offers a mesoscopic description of the material's structural arrangement~\cite {bird1987dynamics}.
Different numerical methods, like finite volume~\cite{owens2002computational}, finite element~\cite{davies1984numerical,fan1999galerkin}, and finite difference methods have been used to solve these equations. Nevertheless, when the Weissenberg number is large, representing the ratio between the elastic and viscous forces, the challenge shifts to ensuring accuracy and stability of the numerical scheme. One way to tackle this is rather than solving the conformation tensor directly, to instead compute the evolution of the logarithm of the conformation tensor. This computational methodology ensures the positive-definite nature of the conformation tensor and provides a promising avenue to compute viscoelastic flows~\cite{fattal2004constitutive,fattal2005time,hao2007simulation}. 

Experimental investigations of the coalescence of viscoelastic droplets have shown that compared to Newtonian drop coalescence, viscoelastic droplets exhibit the formation of sharper bridges~\cite{dekker2022elasticity, bouillant2022rapid}. Accurate interface capturing and curvature estimation in numerical studies is then essential to predict such flows. Different ways to do this include the volume of fluid~\cite{scardovelli1999direct,popinet2003gerris} and the level set method~\cite{sussman1994level,osher2001level}, which are considered as sharp interface methods, while the phase-field method is the diffused interface method~\cite{jacqmin1999calculation,yue2004diffuse,geier2015conservative,lee2010lattice}, i.e. the interface has a finite thickness. More recently, the conservative phase-field method ~\cite{sun2007sharp,chiu2011conservative} is developed to solve the mass loss problem when simulating small droplets~\cite{yue2007spontaneous,zheng2014shrinkage}, which largely improves the efficiency of the $4^{th}$ order Cahn-Hilliard equation. In our previous work, the three-phase conservative phase-field lattice Boltzmann method was applied to compute the Newtonian droplet dynamics generated as it meets a perfectly spreading fluid film/pool. In this study, we extend our computational approach by solving the Navier-Stokes equations with a velocity-pressure lattice Boltzmann method~\cite{zhao2023interaction}. Furthermore, the interface is modeled by the conservative phase-field lattice Boltzmann method~\cite{zhao2023interaction,geier2015conservative} and the surface tension is represented by the continuum surface force model~\cite{brackbill1992continuum,kim2005continuous}. The curvature is estimated by the $2^{nd}$ order isotropic finite difference method, which highly improves the numerical performance as introduced in~\cite{lee2005stable}. Below, we will show the 2D simulations of the three-phase flow when a Newtonian droplet comes in contact an immicible viscoelastic fluid film. 

\section{Computational methodology}\label{methodology}
\begin{figure*}
\centering
  \includegraphics[width=0.8\linewidth]{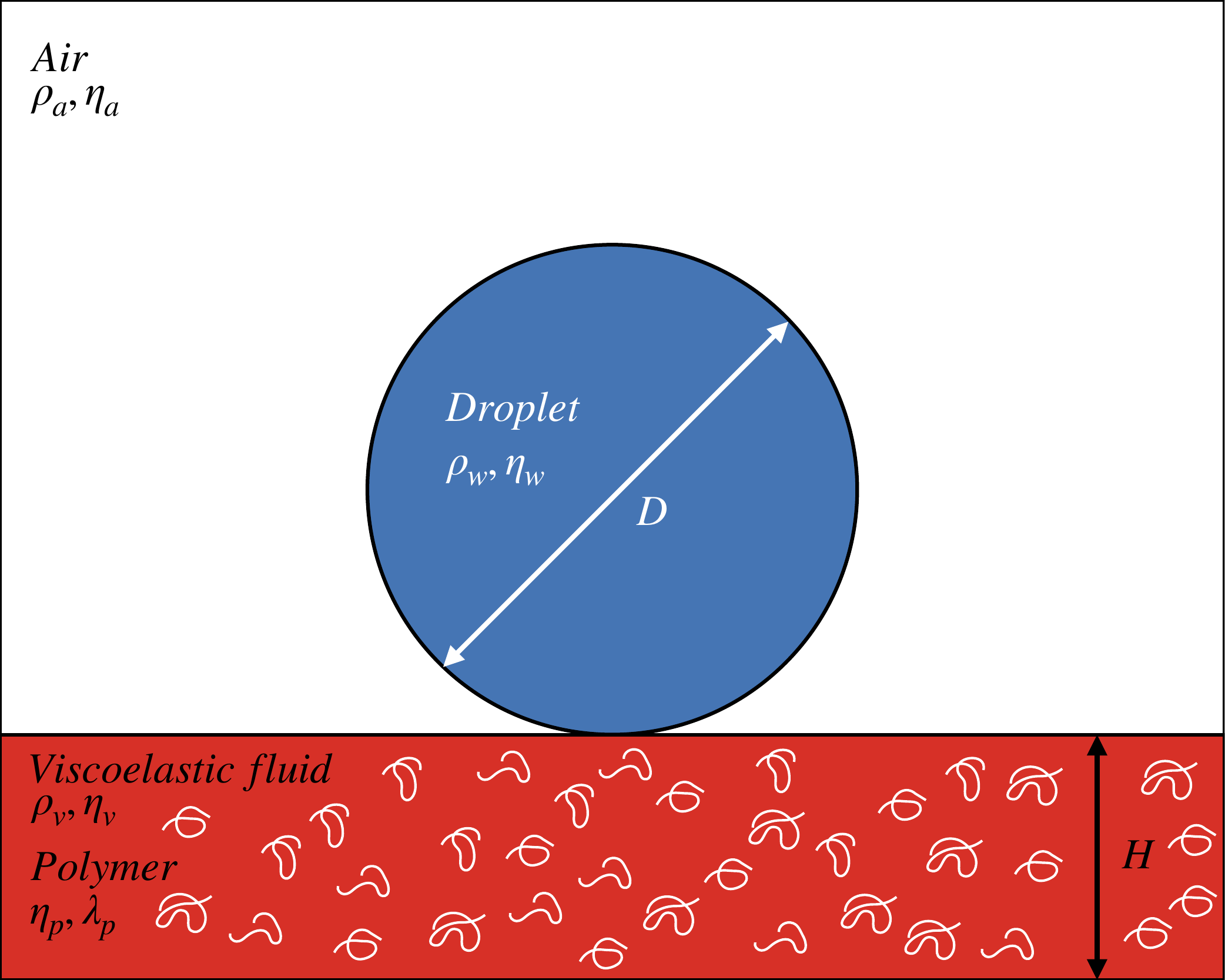}
    \caption{Schematic of the initial condition of the three-phase fluid flow simulations. The dark blue phase denotes the droplet with density $\rho_w$, viscosity $\eta_w$, and diameter $D$. The red phase represents the viscoelastic fluid solvent with density $\rho_v$, viscosity $\eta_v$, and height $H$, which is filled with polymers of viscosity $\eta_p$, and relaxation time $\lambda_p$. The colorless background represents the surrounding air with density $\rho_a$ and viscosity $\eta_a$. \label{initialization} 
    }
\end{figure*}
\subsection{Simulation setup}
 \begin{table}
\caption{\label{Tab1} Physical properties of the liquids used in the ternary flow system.}
\centering
\begin{tabular}{ccc}
 Parameters &~~~~~ Description& Value \\ 
 \hline
    $\rho_v/\rho_a$  &~~~~~ density ratio      ~~~~~& 100    \\
    $\rho_v/\rho_w$  &~~~~~ density ratio    ~~~~~& 1    \\
    $\eta_v/\eta_a$  &~~~~~ viscosity ratio    ~~~~~& 1000    \\
    $\eta_v/\eta_w$  &~~~~~ viscosity ratio ~~~~~& 10    \\
    $\eta_p/\eta_v$  &~~~~~ viscosity ratio ~~~~~& 1-200    \\
    $Cn=\delta/D$    &~~~~~~~~ Cahn number ~~~~~~~ &0.02\\
    $De=\lambda_p/t_0$             &~~~~~~~~ Deborah number                     ~~~~~& $0.025-250$\\
    $H^*=H/D$            &~~~~~ initial film height to drop radius ratio         ~~~~~& $0.05-2$    \\
    $Oh=\eta_v/\sqrt{\rho_v \sigma_{wv}D}$             &~~~~~ Ohnesorge number                     ~~~~~& $0.05,3.7$\\
    $S_v^*=(\sigma_{wa}-\sigma_{wv}-\sigma_{va})/\sigma_{wv}$             &~~~~~ Spreading factor           ~~~~~& $-1.5-1.5$            \\

\end{tabular}

\end{table}
Simulations of the droplet dynamics on a viscoelastic fluid film/pool are conducted on a rectangular domain, see Figure~\ref{initialization}. Initially, we place the droplet with density $\rho_w$, viscosity $\eta_w$, and diameter $D$ on the viscoelastic fluid film/pool with density $\rho_v$, viscosity $\eta_v$ and height $H$, filled with a polymer of viscosity $\eta_p$ and with a relaxation time $\lambda_p$. The initial vertical position of the droplet's center is set as $c_i=H+0.5D+0.01\delta$, where $\delta$ is the interface thickness, a nonphysical simulation parameter of the general diffused interface method~\cite{jacqmin1999calculation}. The background fluid is composed of air with density $\rho_a$, and viscosity $\eta_a$. 

In all simulations, the density and viscosity ratios are fixed as $\rho_w/\rho_a=\rho_v/\rho_a=100$; $\eta_v/\eta_a=1000$, $\eta_w/\eta_a=100$. The Ohnesorge number $Oh=\eta_v/\sqrt{\rho_v \sigma_{wv} D}$ is used to relate the viscous force to the inertia-capillary force. Normally, $Oh\ll 1$ is considered an inertial regime, and $Oh\gg 1$ is considered a viscous regime. To characterize the surface tension effect, we introduce the scaled spreading factor of the viscoelastic liquid  $S_v^*=(\sigma_{wa}-\sigma_{wv}-\sigma_{va})/\sigma_{wv}$. If the spreading factor is positive $S_v^*>0$ when the droplet is placed on the liquid film, the surface tension force will drive the liquid in the film will gradually cover the droplet. In addition, the height effect is characterized by the initial scaled height of the liquid film, denoted by $H^*=H/D$. 

We further introduce the polymer into the liquid film/pool, which introduces two important parameters: the viscosity of the polymer $\eta_p$, and the relaxation time for the polymer $\lambda_p$ are characterized by the viscosity ratio $\eta^*=\eta_p/\eta_v$, and the Deborah number $De=\lambda_p/t_0$ that is the ratio of the polymer relaxation time and the characteristic flow time scale $t_0$. When $Oh<1$, $t_0=t_\rho=\sqrt{\rho_v D^3/\sigma_{wv}}$, while $t_0=t_\eta=\eta_v D/\sigma_{wv}$ for $Oh>1$. We summarize all of the parameter definitions and the values explored in the numerical simulations in Tab.~\ref{Tab1}

\subsection{Governing equations}
The mathematical description of the three-phase viscoelastic flow includes the Navier-Stokes equations, the three-phase conservative phase-field equations, and the conformation equation. Those governing equations are listed below:

\begin{equation}
\frac{\partial \bar{p}}{\partial t}+\boldsymbol{u}\cdot\nabla \bar{p}+ c_s^2 \nabla\cdot \boldsymbol{u}=0,
\label{gneq:1}
\end{equation}

\begin{equation}
\frac{\partial \boldsymbol{u}}{\partial t}+\nabla\cdot(\boldsymbol{uu})=-\frac{1}{\rho}\nabla p+\frac{1}{\rho}\nabla\cdot \eta\left(\nabla\boldsymbol{u}+(\nabla\boldsymbol{u})^T\right)-\frac{3\delta}{2\rho}\sum_{i=1}^{3}\sigma_i\nabla\cdot\left(\frac{\nabla\phi_i}{|\nabla\phi_i|}\right)   |\nabla\phi_i| \nabla\phi_i+\frac{1}{\rho}\nabla\cdot \boldsymbol{\Pi},
\label{gneq:2}
\end{equation}

\begin{equation}
 \frac{\partial\phi_i}{\partial t}+\nabla\cdot(\phi_i\boldsymbol{u})= \nabla\cdot M\left(\nabla\phi_i-\frac{4}{\delta}\frac{\nabla\phi_i}{|\nabla\phi_i|}\phi_i(1-\phi_i)
    +\frac{\phi_i^2}{\sum_{j=1}^3\phi_j^2}\sum_{j=1}^3\frac{4}{\delta}\frac{\nabla\phi_j}{|\nabla\phi_j|}\phi_j(1-\phi_j)\right),
\label{gneq:3}
\end{equation}

\begin{equation}
    \boldsymbol{\Pi}=\frac{\eta_p}{\lambda_p}(\boldsymbol{A}-\boldsymbol{I}),
\label{gneq:4}
\end{equation}
\begin{equation}
    \frac{\partial \boldsymbol{A}}{\partial t}+\left(\boldsymbol{u}\cdot \nabla\right)\boldsymbol{A}=\boldsymbol{A}\left(\nabla\boldsymbol{u}\right)+\left(\nabla\boldsymbol{u}\right)^T\boldsymbol{A}+\frac{1}{\lambda_p}\left(\boldsymbol{A}-\boldsymbol{I}\right).
\label{gneq:5}
\end{equation}
In the pressure evolution equation, Eq.~\ref{gneq:1}, $\bar{p}=p/\rho$, and $p$ denotes the pressure, $\rho$ represents the locally computed density of the three-phase liquid system. $\boldsymbol{u}$ represents the velocity vector, and $c_s$ is the speed of sound. $\phi_i$ are the liquid phases in the system, represented by the subscript $i\in[1-3]$. In the velocity evolution equation, Eq.~\ref{gneq:2}, $\eta$ is the locally computed viscosity of the three-phase liquid system, and $\delta$ represents the interface thickness \cite{zhao2023engulfment}. The third term on the right-hand side of Eq.~\ref{gneq:3} is known as the continuum surface tension force (CSF)~\cite{brackbill1992continuum,zhao2023interaction} and the modified surface tension for component $i$, $\sigma_i$, is calculated by the surface tension between different components $\sigma_i=\left(\sigma_{ij}+\sigma_{ik}-\sigma_{jk}\right)/2$. Here, we use the three-phase conservative phase-field equation Eq.~\ref{gneq:3} to evolve the order parameter $\phi_i$. In the interface region, we have $0<\phi_i<1$, while in the bulk region of the fluid component $i$, we have $\phi_i=1$. The outside of the fluid component $i$ is denoted as $\phi_i=0$, respectively. $M$ represents the mobility, which is a model parameter. The influence of $M$ and $\delta$ in the conservative phase-field method can be found in \cite{zhao2022ternary, zhao2023engulfment}, and in the current study, we fix $M\Delta t/(\Delta x)^2=0.1$ and $\delta/\Delta x=4$, where $\Delta x $ and $\Delta t$ are lattice units of length and time given as: $\Delta x=1$, $\Delta t=1$. To solve Eqs.~\ref{gneq:1}, \ref{gneq:2}, and \ref{gneq:3}, we employ the velocity-pressure lattice Boltzmann method~\cite{zhao2023interaction}, and the conservative phase-field lattice Boltzmann method~\cite{geier2015conservative,zhao2023interaction}. In our previous study, we have shown that the conservative phase-field lattice Boltzmann method is able to recreate the experimental three-phase water-oil-air interfacial flow~\cite{zhao2023engulfment}. Therefore, we do not repeat the methodology of the numerical schemes for the momentum equations and the phase-field equations, which are reported in~\cite{zhao2023engulfment}. 

The last term of Eq.~\ref{gneq:3} represents the viscoelastic stress  $\boldsymbol{\Pi}$  and can be expressed as Eq.~\ref{gneq:4}. We employ the Oldroyd-B model to solve the viscoelastic stress which can be expressed as Eq.~\ref{gneq:5}. In Eq.~\ref{gneq:4} and \ref{gneq:5}, $\boldsymbol{A}$ is the conformation tensor, and $\boldsymbol{I}$ represents the identity tensor. In our computational method, the time derivative is solved by a $4^{th}$ order Runge-Kutta scheme~\cite{butcher1964implicit}. The derivatives of the different components are evaluated by the isotropic finite difference method~\cite{lee2005stable}. The details of the numerical method for the conformation equation are presented in the appendix.

\section{Simulation Results}\label{simulation}
\subsection{Evolution of the droplet's center on a viscoelastic fluid film}\label{mass}

\begin{figure*}
\centering
  \includegraphics[width=\linewidth]{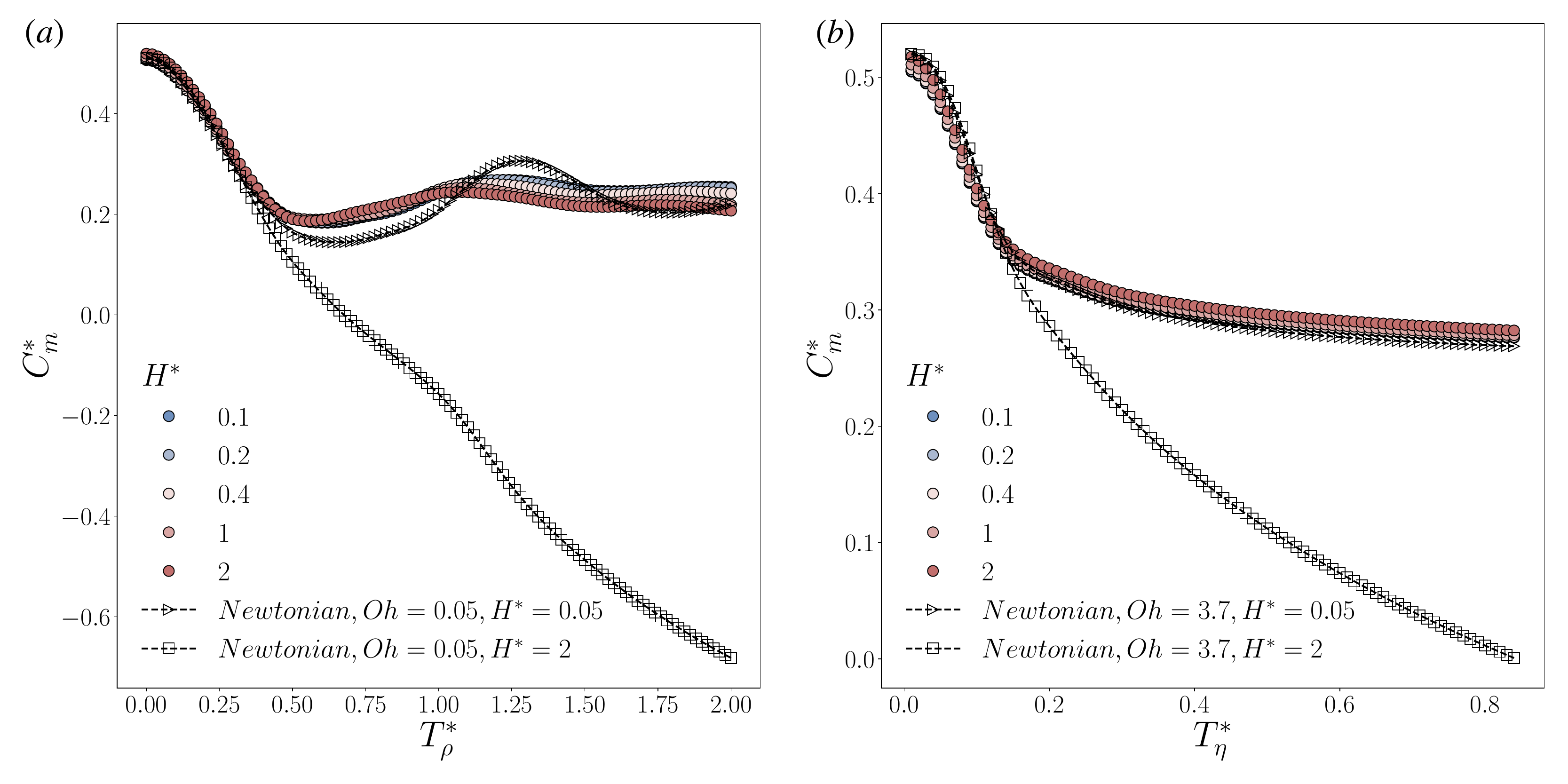}
    \caption{Evolution of the scaled center of mass ($C_m^*$) position (wall normal) of the droplet on viscoelastic fluid film for $De=0.25$, with varying thickness ($H^*=[0.05,2]$) during (a) $T_\rho^*=[0,2.0]$ for inertial regime with $Oh=0.05$, and (b) $T_\eta^*=[0,0.85]$ for viscous regime with $Oh=3.7$. The black dashed lines with triangle markers and square markers indicate the results of the Newtonian flow with liquid heights $H^*=0.05$ and $H^*=2.0$ for each $Oh$ respectively.\label{height1} 
    }
\end{figure*}
\begin{figure*}
\centering
  \includegraphics[width=\linewidth]{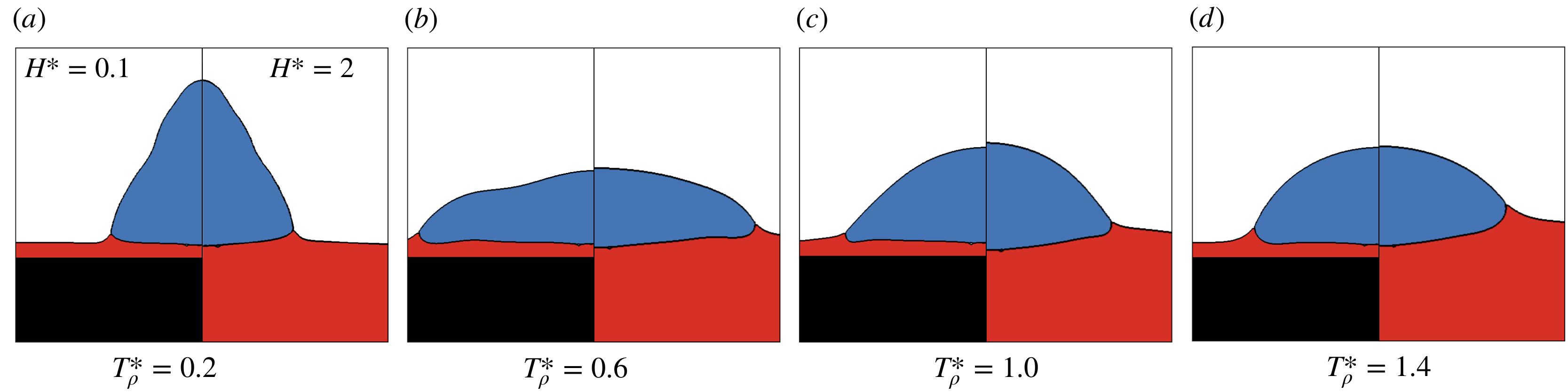}
    \caption{Simulations of the droplet being spreading on viscoelastic fluid film for $De=0.25$, with different heights $H^*=0.1$ (left panel) and $H^*=2.0$ (right panel) during $T_\rho^*=[0.2,1.4]$ for the inertial regime with $Oh=0.05$.\label{height2} 
    }
\end{figure*}
\begin{figure*}
\centering
  \includegraphics[width=\linewidth]{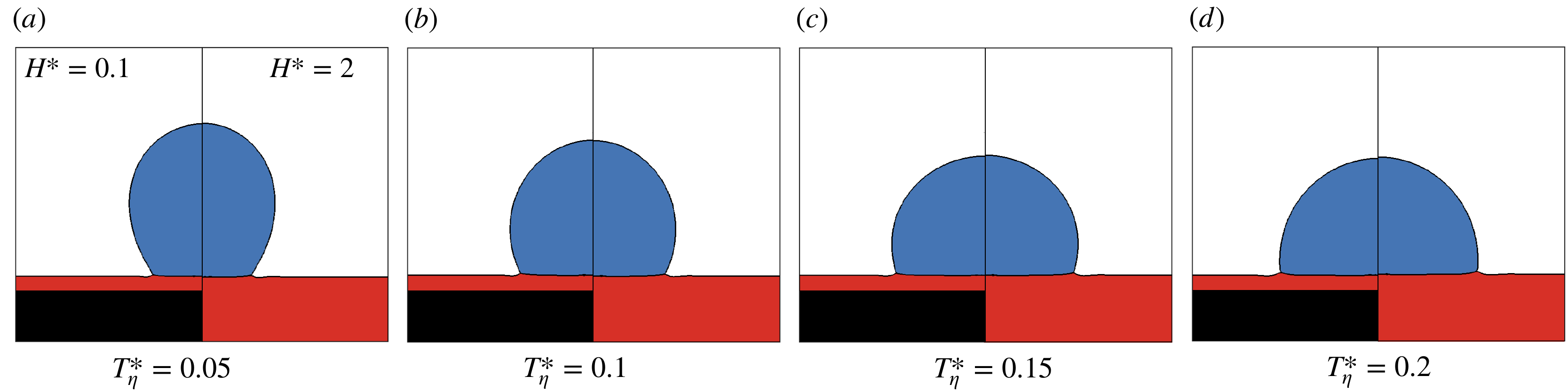}
    \caption{Simulations of the droplet being spreading on viscoelastic fluid film for $De=0.25$, with different heights $H^*=0.1$ (left panel) and $H^*=2.0$ (right panel) during $T_\eta^*=[0.05,0.2]$ for the viscous regime with $Oh=3.7$.\label{height3} 
    }
\end{figure*}
\begin{figure*}
\centering
  \includegraphics[width=\linewidth]{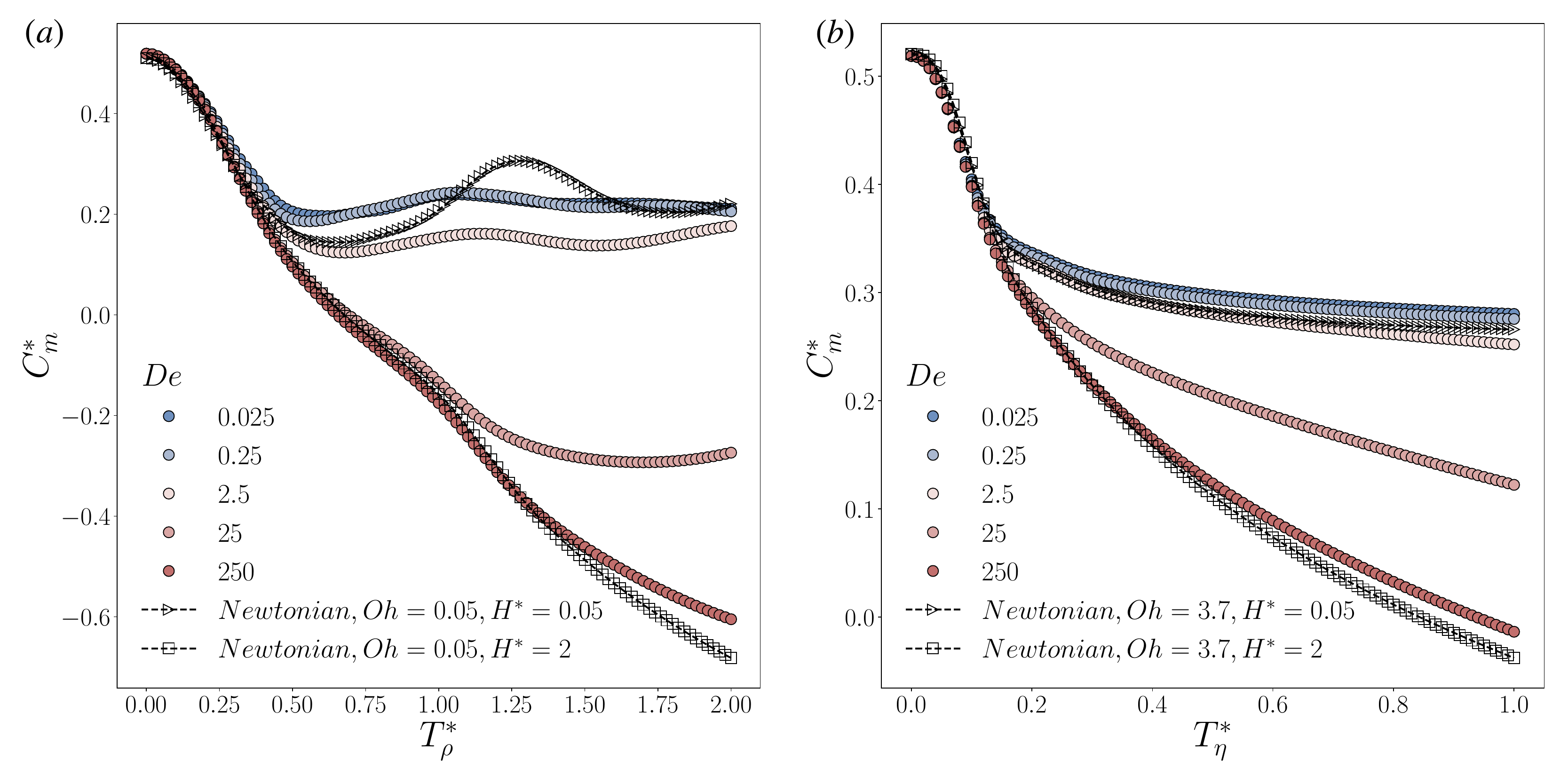}
    \caption{Evolution of the scaled center of mass ($C_m^*$) position (wall normal) of the droplet on viscoelastic fluid pool with varying relaxation time ($De=[0.025,250]$),  during (a) $T_\rho^*=[0,2.0]$ for inertial regime with $Oh=0.05$, and (b) $T_\eta^*=[0,1]$ for viscous regime with $Oh=3.7$. The black dashed lines with triangle markers and square markers indicate the results of the Newtonian flow with liquid heights $H^*=0.05$ and $H^*=2.0$ for each $Oh$ respectively.\label{De} 
    }
\end{figure*}

\begin{figure*}
\centering
  \includegraphics[width=\linewidth]{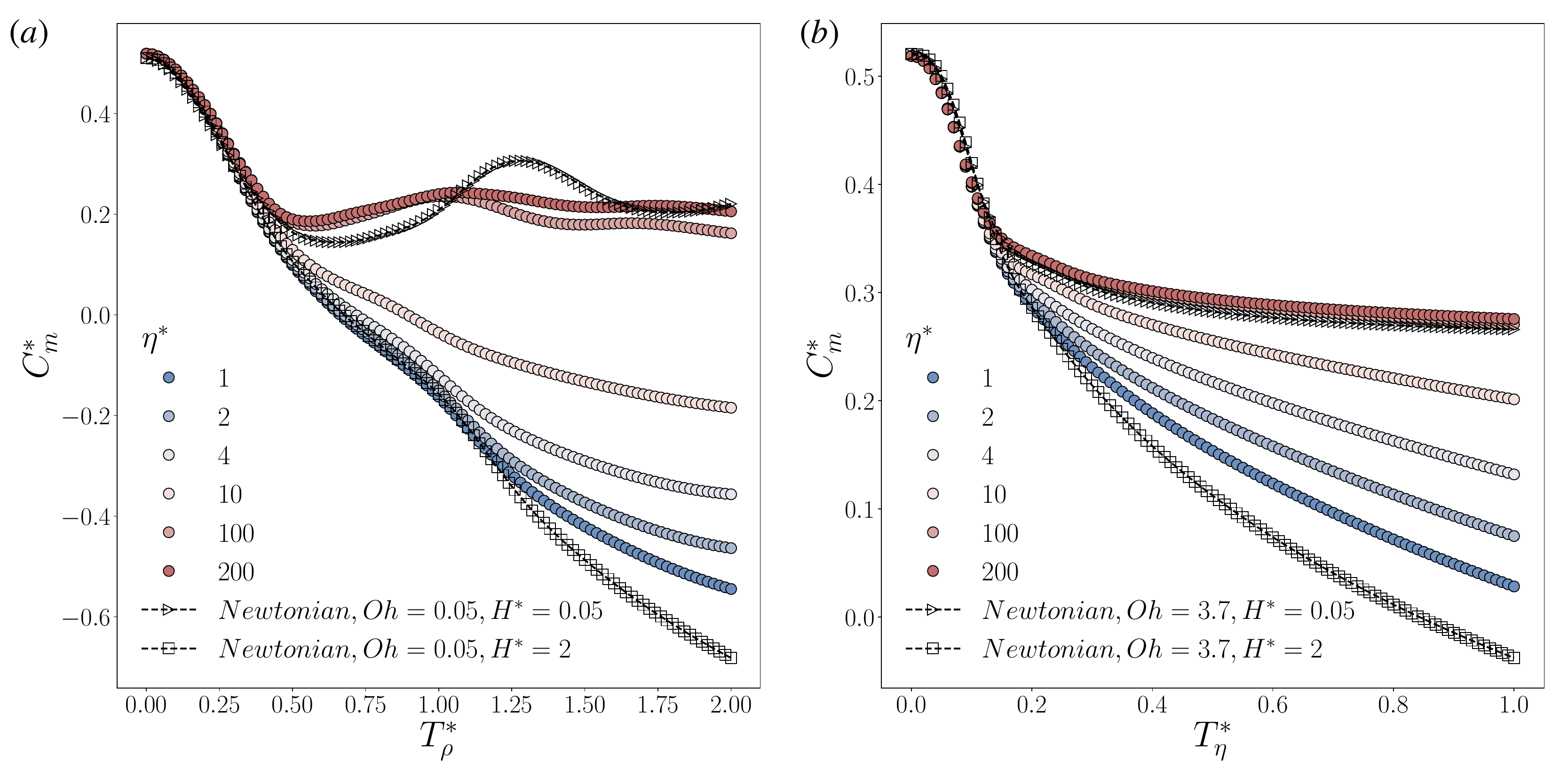}
    \caption{Evolution of the scaled center of mass ($C_m^*$) position (wall normal) of the droplet on viscoelastic fluid pool with varying polymer viscosity ($\eta^*=[1,200]$),  during (a) $T_\rho^*=[0,2.0]$ for inertial regime with $Oh=0.05$, and (b) $T_\eta^*=[0,1]$ for viscous regime with $Oh=3.7$. The black dashed lines with triangle markers and square markers indicate the results of the Newtonian flow with liquid heights $H^*=0.05$ and $H^*=2.0$ for each $Oh$ respectively.\label{vp} 
    }
\end{figure*}

\begin{figure*}
\centering
  \includegraphics[width=\linewidth]{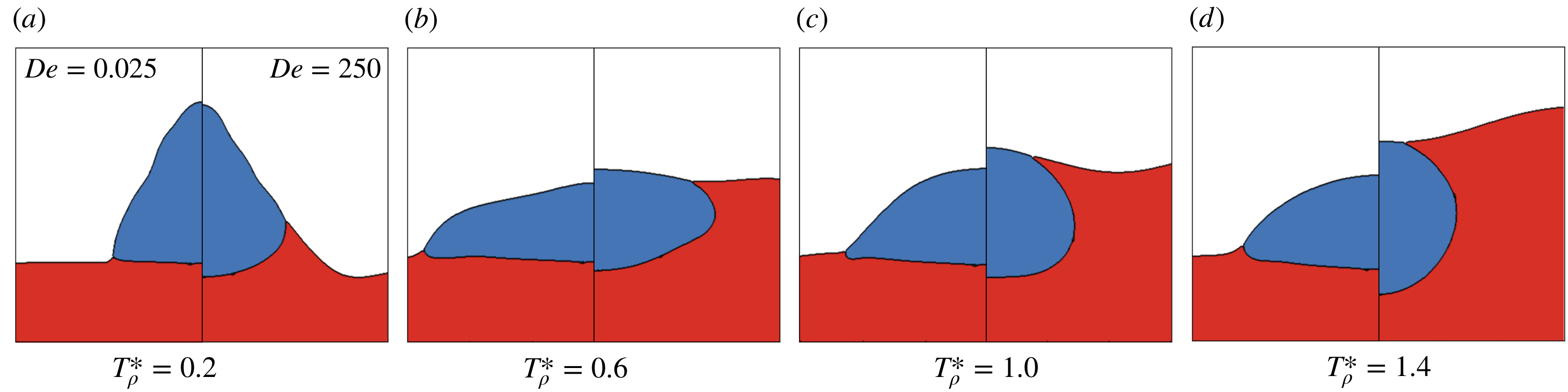}
    \caption{Simulations of the droplet being spreading on a viscoelastic fluid pool with different heights $H^*=0.1$ (left panel) and $H^*=2.0$ (right panel) during $T_\rho^*=[0.2,1.4]$ for the inertial regime with $Oh=0.05$.\label{inde} 
    }
\end{figure*}
\begin{figure*}
\centering
  \includegraphics[width=\linewidth]{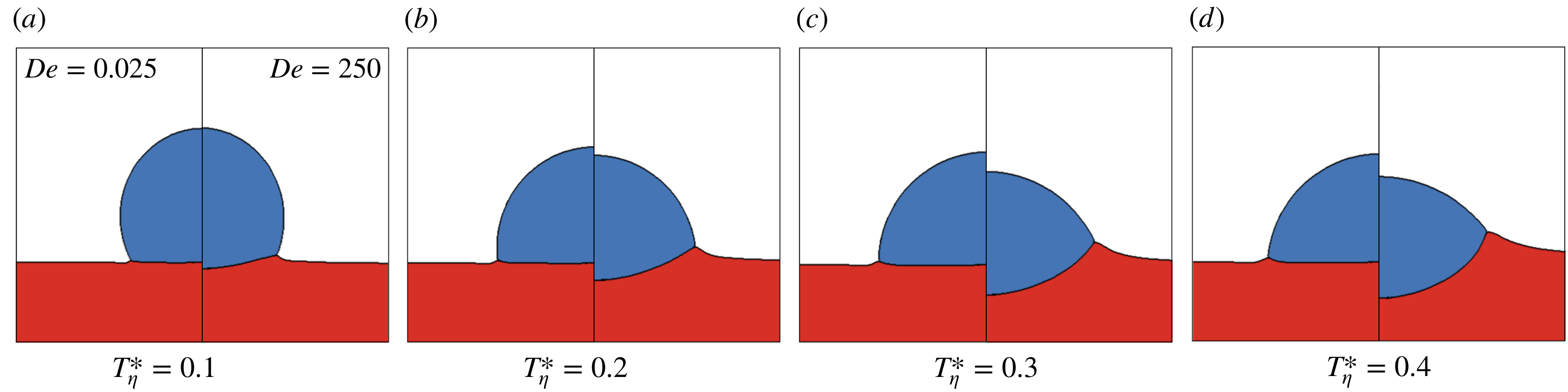}
    \caption{Simulations of the droplet being spreading on viscoelastic fluid pool with different heights $H^*=0.1$ (left panel) and $H^*=2.0$ (right panel) during $T_\eta^*=[0.05,0.2]$ for the viscous regime with $Oh=3.7$.\label{visde} 
    }
\end{figure*}
\subsubsection{Effect of the height of the viscoelastic fluid film}
We start to systematically vary the dimensionless film height of the pre-coated solid, denoted as $H^*$, within both the inertial and viscous regimes. In these numerical experiments, the Deborah number $De=0.25$ and the viscosity ratio $\eta^*=200$ are fixed. Figure~\ref{height1} (a) illustrates the evolution of the center of mass  $C_m^*=C_m/D$ with time for different $H^*$ values in the inertial regime with an Ohnesorge number $Oh=0.05$. To help contrast the Newtonian behaviour, we include two dashed reference curves with markers representing  Newtonian flow with $Oh=0.05$, corresponding to $H^*=0.1,2$. There are some noteworthy points to highlight about the behaviour of the viscoelastic system. It appears that the dynamics for $De=0.25$ are fairly insensitive to the relative height $H^*$. It can be seen that the final position of the center of mass is close to $C_m^*=0.2$, similar to the Newtonian flow with $H^*=0.05$. However, the viscoelastic nature of the fluid film acts to dampen the inertia-capillary wave that is visible for the Newtonian flow.

In Figure~\ref{height1}(b), we present the results of a series of simulations conducted within the viscous regime, with $Oh=3.7$. The drop dynamics are insensitive to the thickness of the film. Instead, a master curve for $C^*_m$ appears following a very similar behaviour to the dynamics of a Newtonian droplet on a Newtonian thin liquid film.

\subsubsection{Effect of the polymer relaxation time $\lambda_p$}\label{engulf}
The droplet dynamics is found insensitive to the pre-wetted film height when the polymer's viscosity is high. The resulting dynamics are then reminiscent of the Newtonian case for thin pre-wetted film and it appears the viscoelastic film becomes more solid-like.

Next, we determine how $De$ affects the viscoelastic engulfment process. We systematically vary $De$ within the range of $[0.025, 250]$ by adjusting the relaxation time of the polymer, denoted as $\lambda_p$. It is worth noting that we fix the viscosity ratio ($\eta^*$) and the liquid film height ($H^*=2$) throughout these simulations. As demonstrated in Eq.~\ref{gneq:4} and \ref{gneq:5}, the increase in $\lambda_p$ leads to a reduction in the viscoelastic stress. Additionally, the relaxation term introduced in Eq.~\ref{gneq:5}, denoted as $(\boldsymbol{A}-\boldsymbol{I})/\lambda_p$, diminishes as $\lambda_p$ increases.

Figure~\ref{De}(a) illustrates the vertical motion of the droplet's center of mass position for different $De$ for inertially dominated dynamics $Oh=0.05$. The engulfment process exhibits minimal variation when $De$ is within the range of $[0.025, 2.5]$. However, a substantial deviation from this dynamics is observed for large $De=25$. Intriguingly, when $De=250$, the viscoelastic engulfment process closely resembles that of the Newtonian thin film. The Deborah number controls the polymer relaxation time. As we increase the relaxation time, the polymer takes a long time to recover to its original shape which makes it behave like fluid. Therefore, the viscoelastic film with higher Deborah numbers converges to the Newtonian film.

\subsubsection{Effect of the polymer viscosity $\eta_p$ }
We conduct additional numerical simulations to determine the influence of the polymer viscosity on the engulfment process. As indicated in Eq.~\ref{gneq:4}, the polymer viscosity plays an important role in determining the magnitude of the stress induced by the viscoelastic polymer. As we systematically increase the polymer viscosity, it becomes increasingly evident that the stress attributed to the polymer induces pronounced changes to the interfacial dynamics.

In Figure~\ref{vp} (a), we observe the temporal evolution of the mass center within the inertial regime under varying polymer viscosities. Notably, when the polymer viscosity is relatively low, the engulfment process of the viscoelastic film is fairly similar to the flow of a Newtonian fluid film. However, as we progressively increase the polymer viscosity to sufficiently high levels, the behaviour of the liquid film transitions towards that of a solid substrate, exhibiting elastic characteristics. Figure~\ref{vp} (b) demonstrates similar trends within the viscous regime, where an increase in polymer viscosity, the evolution of the mass center gradually converges towards that observed in the case of a Newtonian thin film liquid.

\subsection{Evolution of the contact angle near the contact line}
\begin{figure*}
\centering
  \includegraphics[width=0.8\linewidth]{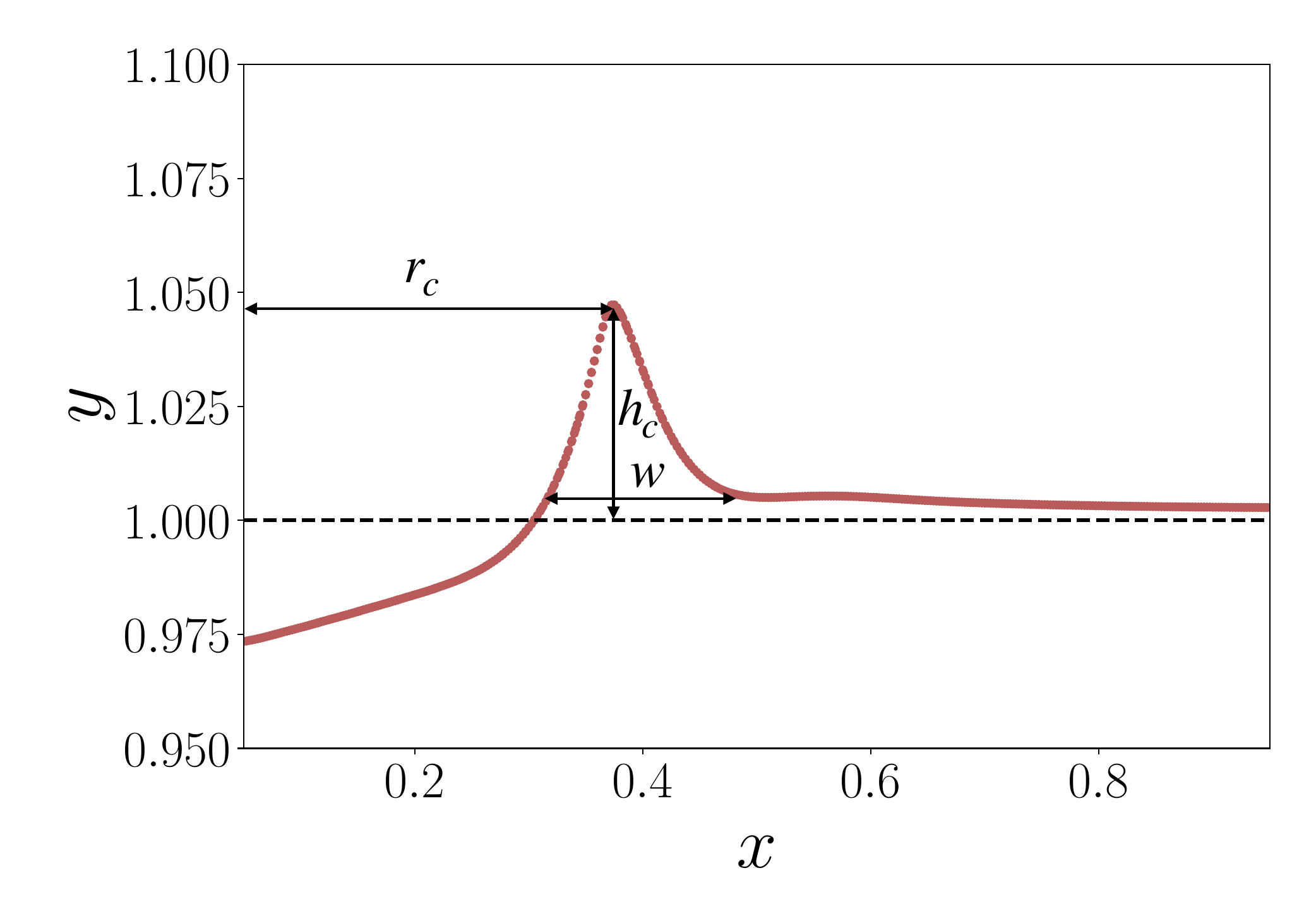}
    \caption{Interface of the liquid film $\phi_v=0.5$ for $G^*=80$, $S_v^*=1.5$, and $Oh=3.7$ when $T_\eta^*=0.6$. The horizontal distance between the contact line to the axis is denoted as $r_c$, the distance between the contact line to the initial liquid film interface (indicated by the black dashed line) is denoted as $h_c$, and the width of the wetting ridge (the horizontal distance between the left or the right minimum point of the contact line and the contact surface on another side) is defined as $w$. \label{interface} 
    }
\end{figure*}
In Section~\ref{mass}, we have established that changing the polymer viscosity ($\eta_p$) and the polymer relaxation time ($\lambda_p$) can result in the emergence of an elastic effect limiting the dynamics. To classify the dynamics we employ the scaled shear modulus $G^*=\eta^*/De$ to combine both the effect of the polymer viscosity and the polymer relaxation time. One particularly interesting aspect to establish is how it affects the wetting ridge at the contact line. Previous studies of the deformation of the thin film near the ridge for viscoelastic solids can be found in~\cite{chan2022growth,leong2020droplet,tamim2023spreading}. It is noted that in contrast to works done in~\cite{leong2020droplet,tamim2023spreading}, our simulations are conducted based on the complete Navier-Stokes equations going beyond the thin film approximation. In addition, we consider the fluid flow inside of the viscoelastic film which highly affects the scaling of the evolution of the meniscus tip with time.

As shown in Figure~\ref{interface}, the red dotted surface represents the interface ($\phi_v=0.5$) when $G^*=80$, $S_v^*=1.5$, and $Oh=3.7$ at $T_\eta^*=0.6$. The contact line, i.e. the point where all phases meet, is marked as the highest vertical point of the interface. The distance from the contact line to the initial interface of the pre-wetted viscoelastic liquid film (indicated by the black dashed line) is denoted as $h_c$, the distance from the contact line to the symmetry axis (y-axis) is denoted as $r_c$, the width of the wetting ridge (the horizontal distance between the left or the right minimum point of the contact line and the contact surface on another side) is defined as $w$. The final shape of the droplet, when placed on a liquid film, is determined by the spreading factor~\cite{zhao2023engulfment}. In this context, the system strives to minimize surface energy. We describe here the dynamics of the formation of the viscoelastic wetting ridge as a function of the spreading factor as well as highlighting the elastic stress distribution.

\subsubsection{Effect of the viscoelastic fluid spreading factor $S_v$}
\begin{figure*}
\centering
  \includegraphics[width=\linewidth]{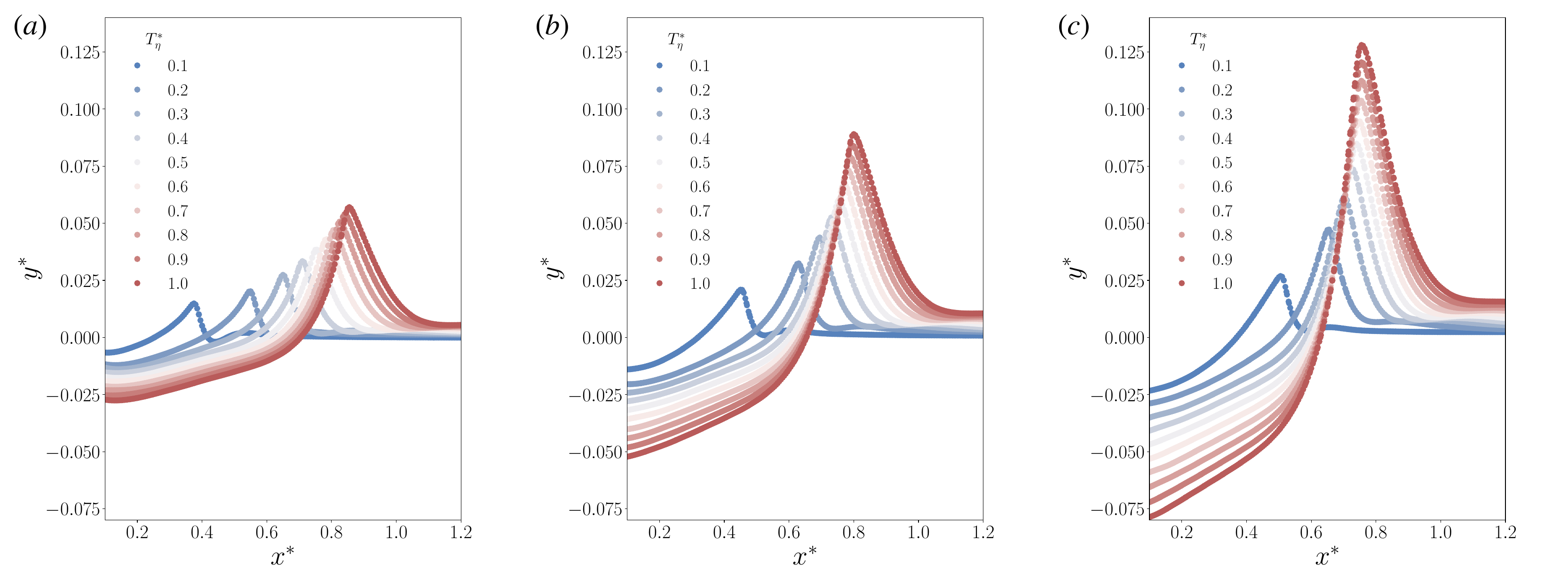}
    \caption{Evolution of the viscoelastic meniscus for varying the scaled spreading factor $S_v^*$, (a) $S_v^*=-1.5$, (b) $S_v^*=0$, and (c) $S_v^*=1.5$, during $T_\eta^*=[0.1,1]$ in the viscous regime with $G^*=80$, $Oh=3.7$. The coordinate is scaled by $x^*=x/D$ and $y^*=(y-H)/D$.\label{spreading1} 
    }
\end{figure*}

\begin{figure*}
\centering
  \includegraphics[width=\linewidth]{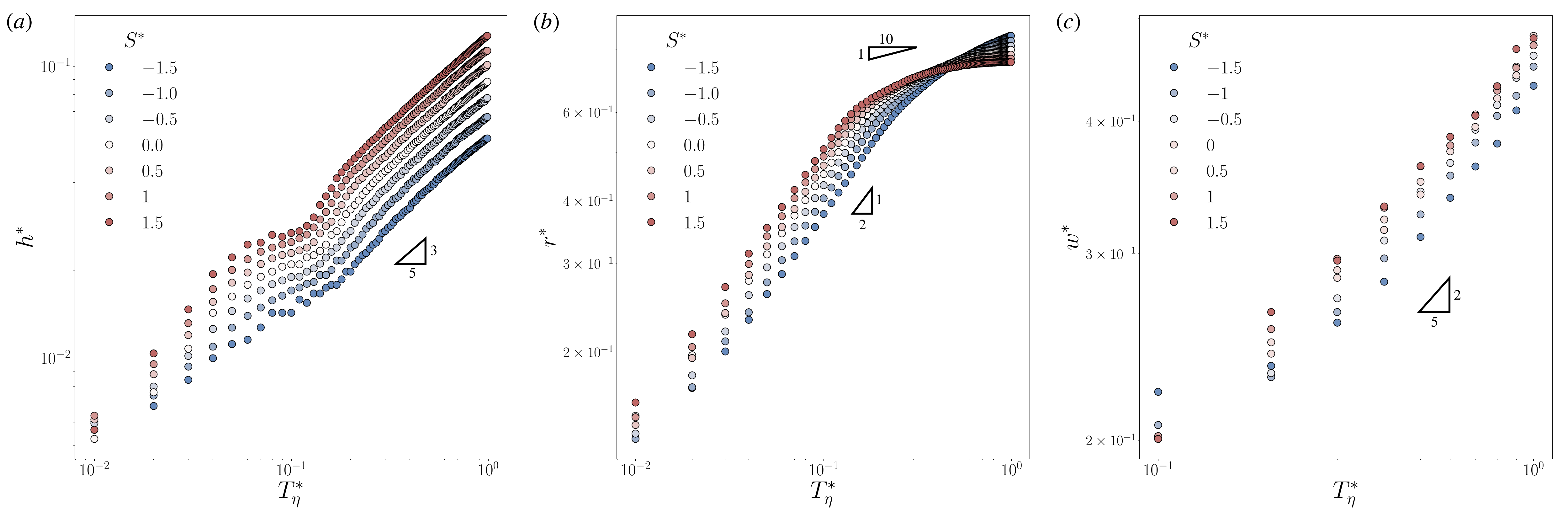}
    \caption{Evolution of (a) the scaled meniscus height $h^*$, (b) the scaled spreading radius $r^*$ at the meniscus tip during $T_\eta^*=[0.01,1]$, and (c) the scaled width of the wetting ridge $w^*$ for varying the scaled spreading factor $S_v^*=[-1.5,1.5]$  in the viscous regime with $G^*=80$,  $Oh=3.7$.\label{spreading2} 
    }
\end{figure*}
In Figure~\ref{spreading1}, we show the evolution of the interface profiles of the viscoelastic wetting ridge around the contact line for different spreading factors: (a) $S_v^*=-1.5$, (b) $S_v^*=0$, and (c) $S_v^*=1.5$ where we have fixed the parameters, $G^*=80$, $H^*=2.0$, and $Oh=3.7$. The spreading factor dictates the magnitude of the deformation of the viscoelastic film, also affecting the aspect ratio of the wetting ridge. 

In Figure~\ref{spreading2}(a), (b), and (c) we present the scaled height, radius, and width of the wetting ridge at the contact line, denoted as $h^*=(h_c-H)/D$, $r^*=r_c/D$, and $w^*=w/D$, respectively, for the contact line. The spreading factor affects the short-time dynamics when $T_\eta^*<0.1$, where $h^*\sim \left(T_\eta^*\right)^\alpha$, and $\alpha$ increases from $0.5$ to $0.75$ as we change $S_v^*=[-1.5,1.5]$. After $T_\eta^*>0.1$, $h^*$ appears to follow the same power-law for different $S_v^*$, $h^*\sim \left(T_\eta^*\right)^{0.6}$, and $S_v^*$ mainly affects its prefactor. 

The behaviour of the spreading radius $r^*$ exhibits a similar power-law relationship with time where $r^*\sim \left(T_\eta^*\right)^\beta$, and $\beta\approx0.5$ for the short-time, $T_\eta^*<0.1$. It is consistent with the observations made during spreading on Newtonian fluid films, as noted in previous research~\cite{zhao2023engulfment}. The lack of a viscoelastic effect is likely a consequence of the shear flow, with little extensional effects on the polymers. Nevertheless, after the short-time spreading, $T_\eta^*>0.1$, the spreading factor starts to affect the spreading radius, and it is observed that the power law exponent changes from $\beta\approx[0.08,0.25]$. 

We further focus on the width of the wetting ridge after its formation  $T_\eta^*>0.1$. As shown in figure~\ref{spreading2}(c), as we modify the spreading factor $S^*=[-1.5,1.5]$, the width of the wetting ridge of different spreading factors follows a similar power law where $w^*\sim(T_\eta^*)^{0.4}$. The results indicate the width of the wetting ridge is highly related to the shear modulus but not the spreading factors after short time evolution. 

\subsubsection{Effect of the shear modulus $G^*$}

\begin{figure*}
\centering
  \includegraphics[width=\linewidth]{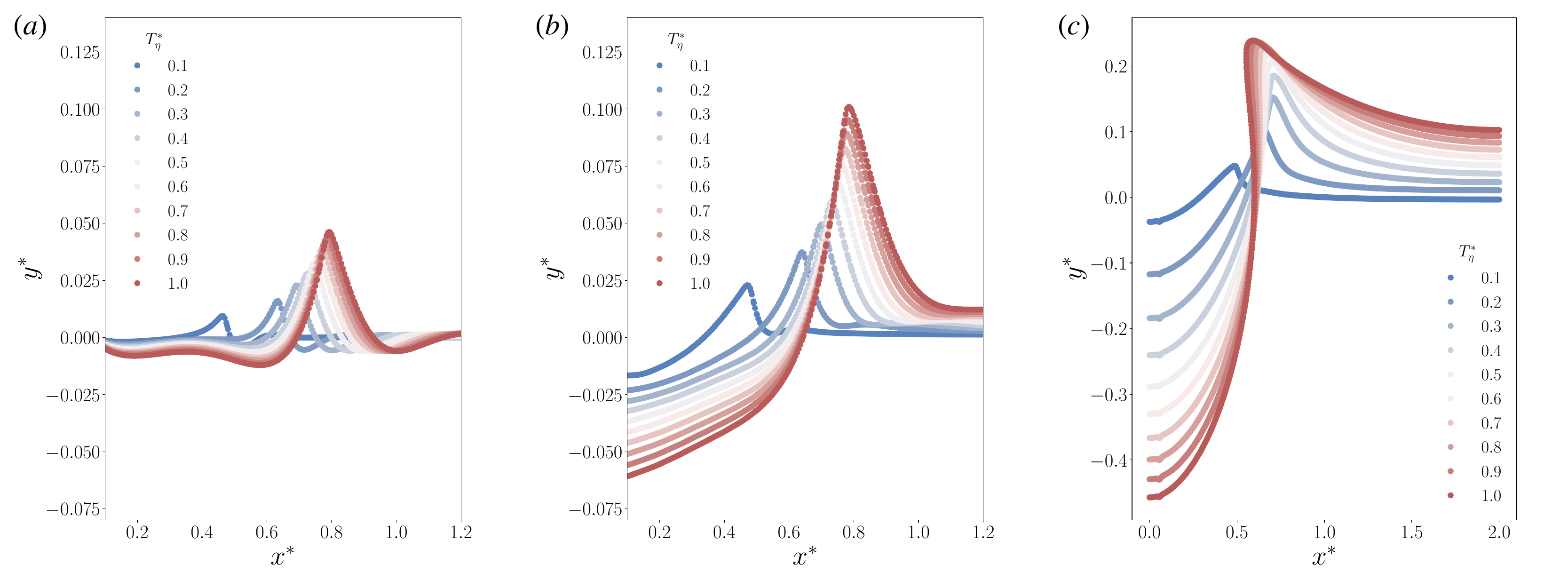}
    \caption{Evolution of the contact lines between the viscoelastic fluid pool with the air and droplet for varying the scaled shear modulus $G^*$, (a) $G^*=8000$, (b) $G^*=80$, and (c) $G^*=0.8$, during $T_\eta^*=[0.1,1]$ in the viscous regime with $S_v^*=0.5$, $Oh=3.7$. The coordinate is scaled by $x^*=x/D$ and $y^*=(y-H)/D$.\label{de1} 
    }
\end{figure*}

\begin{figure*}
\centering
  \includegraphics[width=\linewidth]{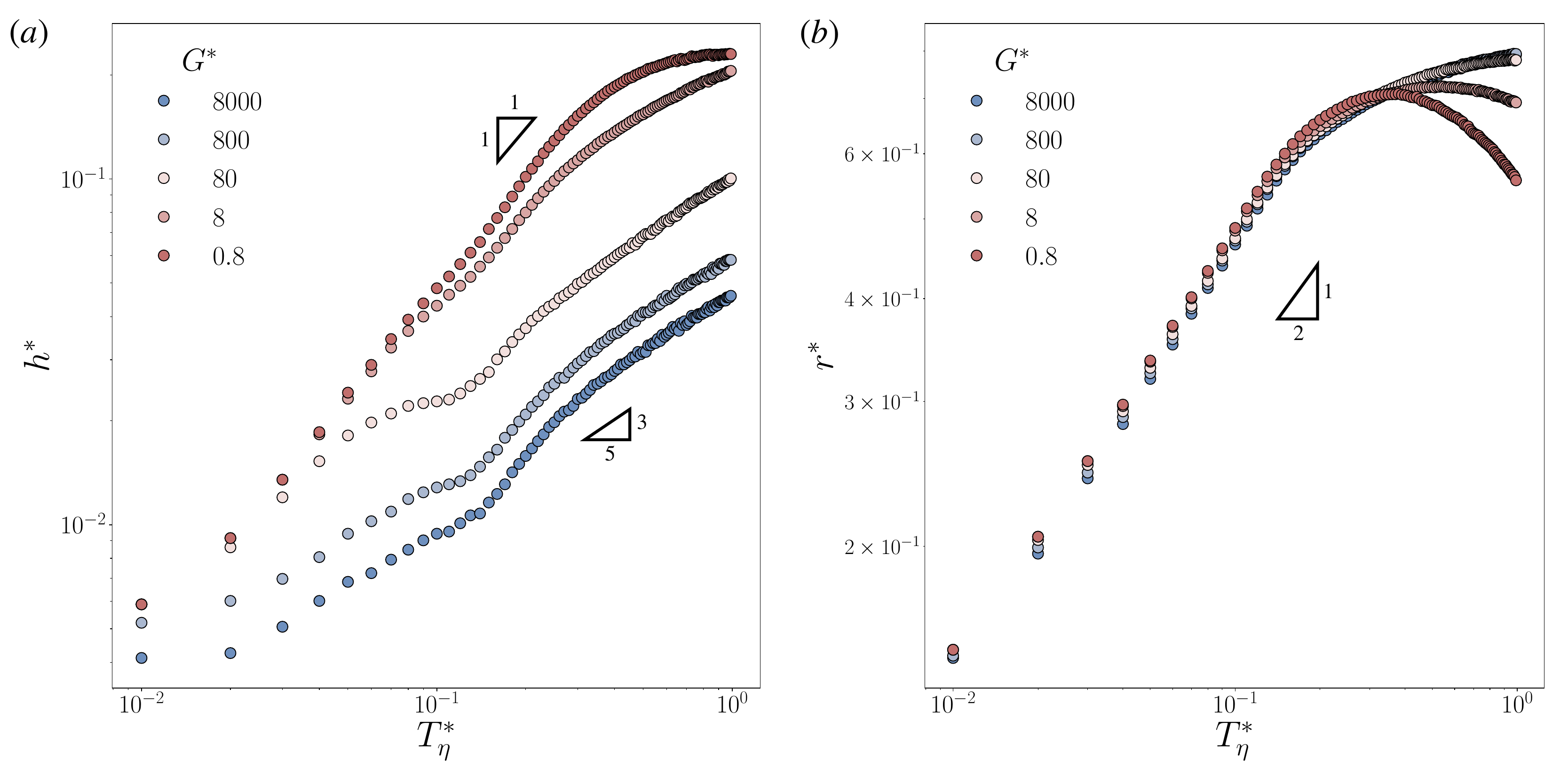}
    \caption{Evolution of (a) the scaled relevant height $h^*$, and (b)the scaled spreading radius $r^*$ of the contact angle tip for varying the scaled spreading factor $G^*=[0.8,8000]$ during $T_\eta^*=[0.01,1]$ in the viscous regime with $S^*=0.5$, $Oh=3.7$.\label{de2} 
    }
\end{figure*}
We proceed to explore the influence of polymer viscosity on the wetting ridge dynamics. Throughout these simulations, we fix $S_v^*=0.5$ as the spreading factor only shifts the data with a pre-factor, while systematically changing $G^*$, within the range of $[0.8,8000]$ by adjusting the relaxation time, $\lambda_p$.

In Figure~\ref{de1} (a), (b), and (c), we present the evolution of the interface near the contact line for $G^*=[0.8,8000]$. As seen in the results, the contact line dynamics can be affected by $G^*$. As we gradually increase $G^*$, the height of the wetting ridge increases faster, and the liquid film exhibits increased compliance. In the case of $G^*=0.8$, the interface of the liquid film undergoes significant deformation, ultimately resulting in the complete engulfment of the droplet by the liquid film.

In Figure~\ref{de2}(a) and (b), we plot the temporal evolution of $h^*$ and $r^*$ in logarithmic axis. Notably, the vertical ridge position exhibits a behaviour affected by $G^*$. For a large $G^*=8000$, the evolution of the vertical position can be separated into two distinct regimes. A short time regime, when $T_\eta^*<0.1$, where it appears that $h^*\sim \left(T_\eta^*\right)^{0.4}$, followed by $h^*\sim \left(T_\eta^*\right)^{0.6}$. In contrast, when we decrease $G^*=0.8$, the evolution of $h^*$ follows a nearly straight line $h^*\sim T_\eta^*$. A similar observation is seen in~\cite{leong2020droplet}, i.e., a linear slope $h^*\sim T_\eta^*$, can be obtained for different shear modulus, while the initial spreading stage was not explored in their work. In addition, the coalescence of two droplets will induce different surface effects which accounts for the difference between their works and our results.

The horizontal position $r^*$ is contrary to $h^*$ insensitive to $G^*$, where $r^*\sim \left(T_\eta^*\right)^{0.5}$. The effect of $G^*$ only affects the late-time dynamics associated with the viscoelastic film engulfing the droplet. In the case of small $G^*$, the contact line moves across the entire droplet's interface, resulting in a reduction in radius. Conversely, for large $G^*$, the spreading process continues but at a much slower rate.

As for the width of the wetting ridge, when we increase the shear modulus $G^*=[80,8000]$, the power law does not change a lot which follows $w^*\sim\left(T_\eta^*\right)^{0.4}$. While, as we decrease $G^*$, which lowers the viscoelastic effect, the wetting ridge is not able to be formed. Therefore, it does not follow any power law under such a low shear modulus.  


\subsubsection{Relaxation of the contact angle}

\begin{figure*}
\centering
      \includegraphics[width=\linewidth]{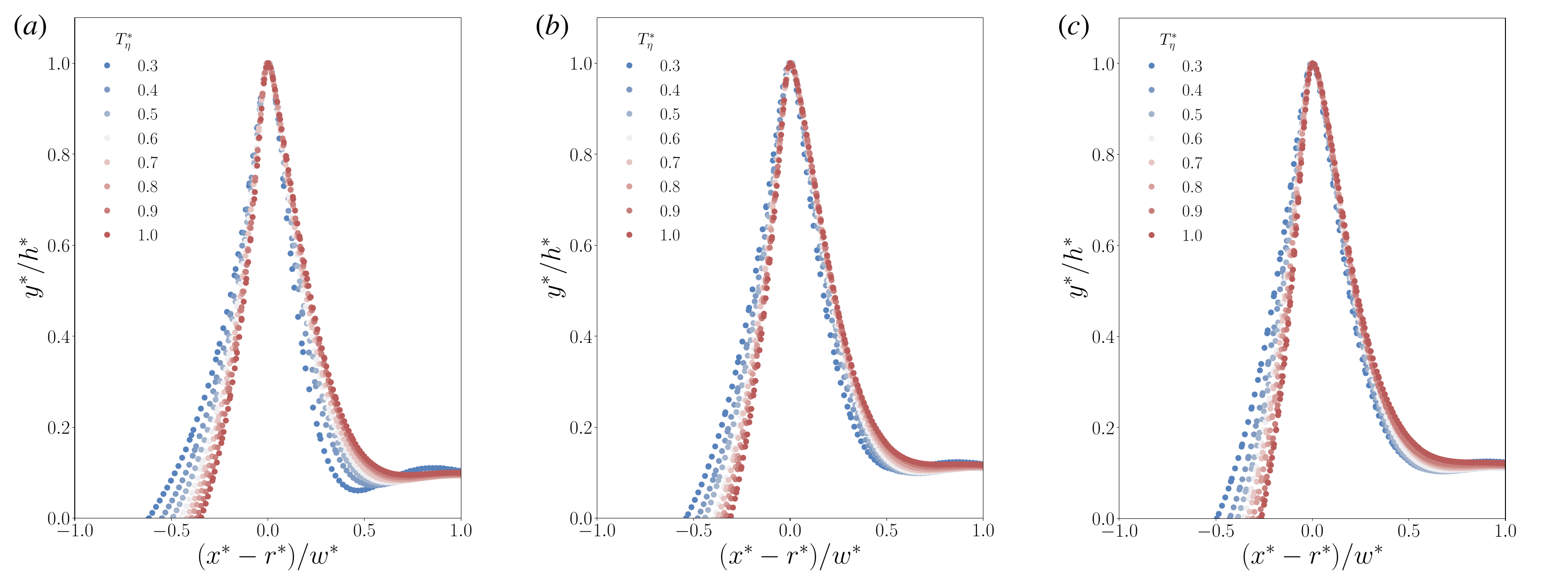}
    \caption{A seemingly universal shape of the wetting ridge appear when recalling the data with the width and height of the ridge for different scaled spreading factor $S_v^*$, (a) $S_v^*=-1.5$, (b) $S_v^*=0$, and (c) $S_v^*=1.5$, during $T_\eta^*=[0.3,1]$ in the viscous regime with $G^*=80$, $Oh=3.7$.\label{relax1} 
    }
\end{figure*}

\begin{figure*}
\centering
  \includegraphics[width=0.66\linewidth]{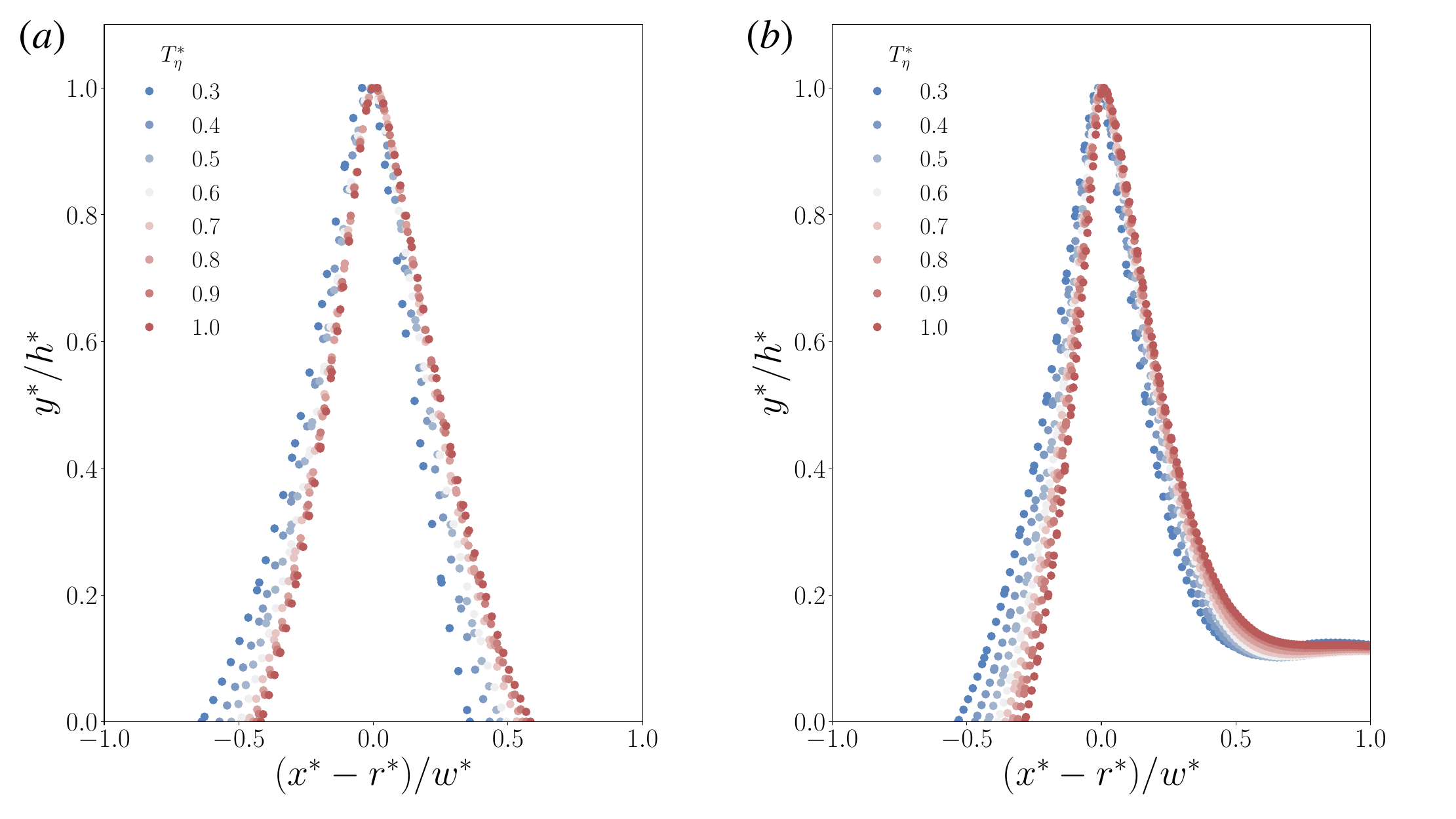}
    \caption{A seemingly universal shape of the wetting ridge appear when recalling the data with the width and height of the ridge for different $G^*$, (a) $G^*=8000$, and (b) $G^*=80$, during $T_\eta^*=[0.3,1]$ in the viscous regime with $S^*_v=0.5$, $Oh=3.7$.\label{relax2} 
    }
\end{figure*}

Figures~\ref{relax1}, \ref{relax2} illustrate what appears to be a self-similar shape of the wetting ridge for parts of our data set. We have here varied both $S_v^*$ and $G^*$, and focus on the late time dynamics i.e., $T_{\eta}> \lambda_p$. The wetting ridge of the viscoelastic film is scaled by $h^*$ and $w^*$.

First, when we fix $G^*=80$ while changing $S_v^*$, we observe that the contact line associated with different $S_v^*$ values closely coincide with one another. Specifically, we observe that the radius scaling power $\beta$ changes from $\beta\sim [0.08,0.23]$ which is consistent with the scaling introduced in the previous section. In addition, the vertical position $h^*$ follows a similar power law where $h^*\sim\left(T^*_\eta\right)^{0.6}$. However, a slight counterclockwise rotation is discerned, attributed to the unbalanced surface tension effects.

Second, by fixing $S_v^*=0.5$ and changing $G^*=[
0.8,8000]$, we observe substantial variations in the interfaces of different simulations. Notably, when $G^*\gg1$, the interface exhibits minimal rotation compared to the short-time profile. The apparent radius of the contact line $r^*\sim \left(T_\eta^*\right)^{0.12}$, the vertical position of the contact line $h^*\sim \left(T_\eta^*\right)^{0.55}$, the width follows $w^*\sim(T_\eta^*)^{0.4}$. Conversely, as $G^*$ decreases, the interface undergoes a more pronounced rotation, and the contact line has a single trend as compared to the cases with higher $G^*$. 

\begin{figure*}
\centering
  \includegraphics[width=\linewidth]{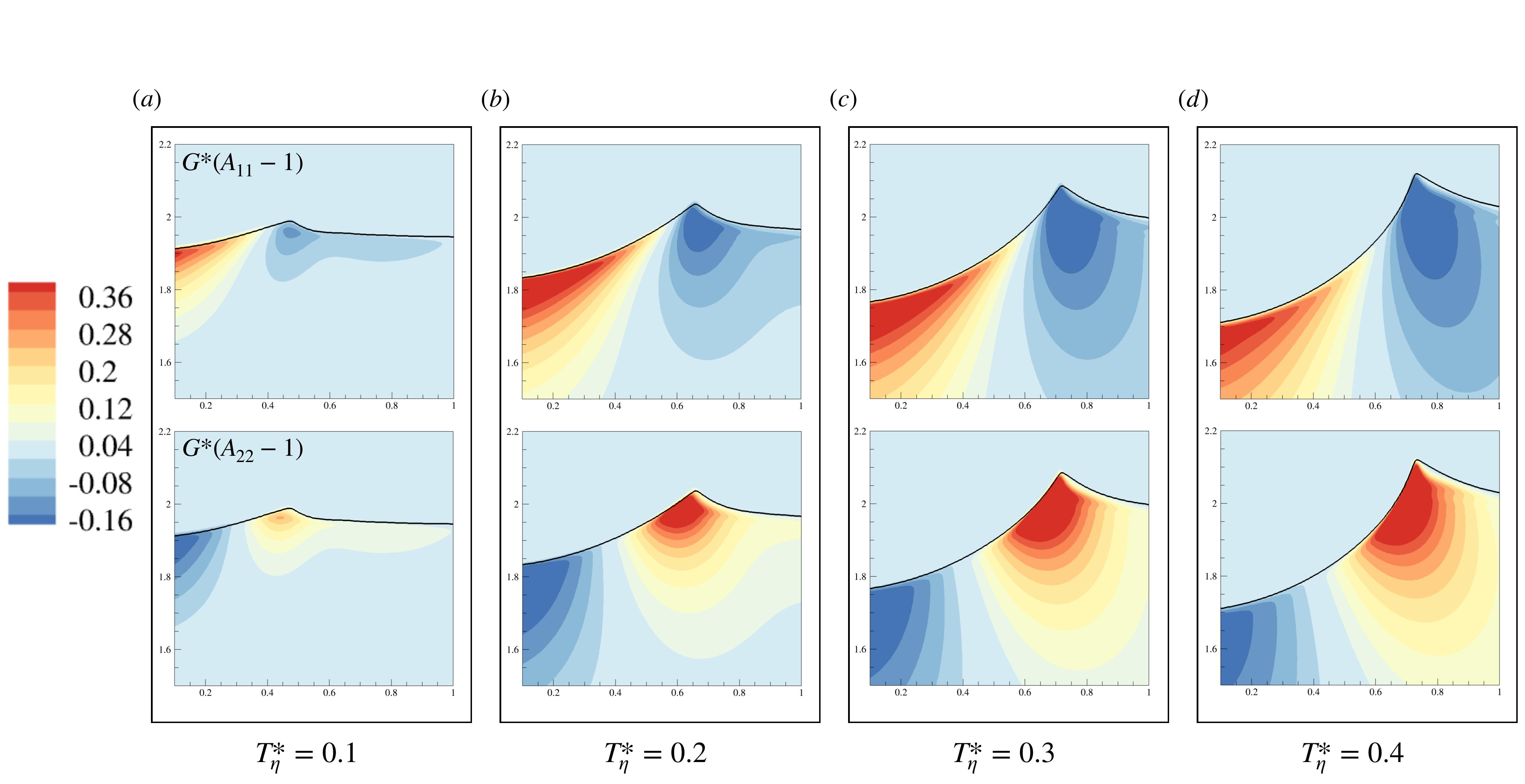}
    \caption{Scaled stress tensor component $\Pi_{11}^*$ (top), and $\Pi_{22}^*$ (bottom) of $\boldsymbol{\Pi}^*=G^*(\boldsymbol{A}-\boldsymbol{I})$ for $G^*=0.8$, $S_v^*=0.5$, $Oh=3.7$,  $T_\eta^*=[0.1,0.4]$. The strain tensor component $A_{11}-1$, and $A_{22}-1$ are in the range of [-0.2,0.45], and an obvious deformation is observed. \label{de1000} 
    }
\end{figure*}

\begin{figure*}
\centering
  \includegraphics[width=\linewidth]{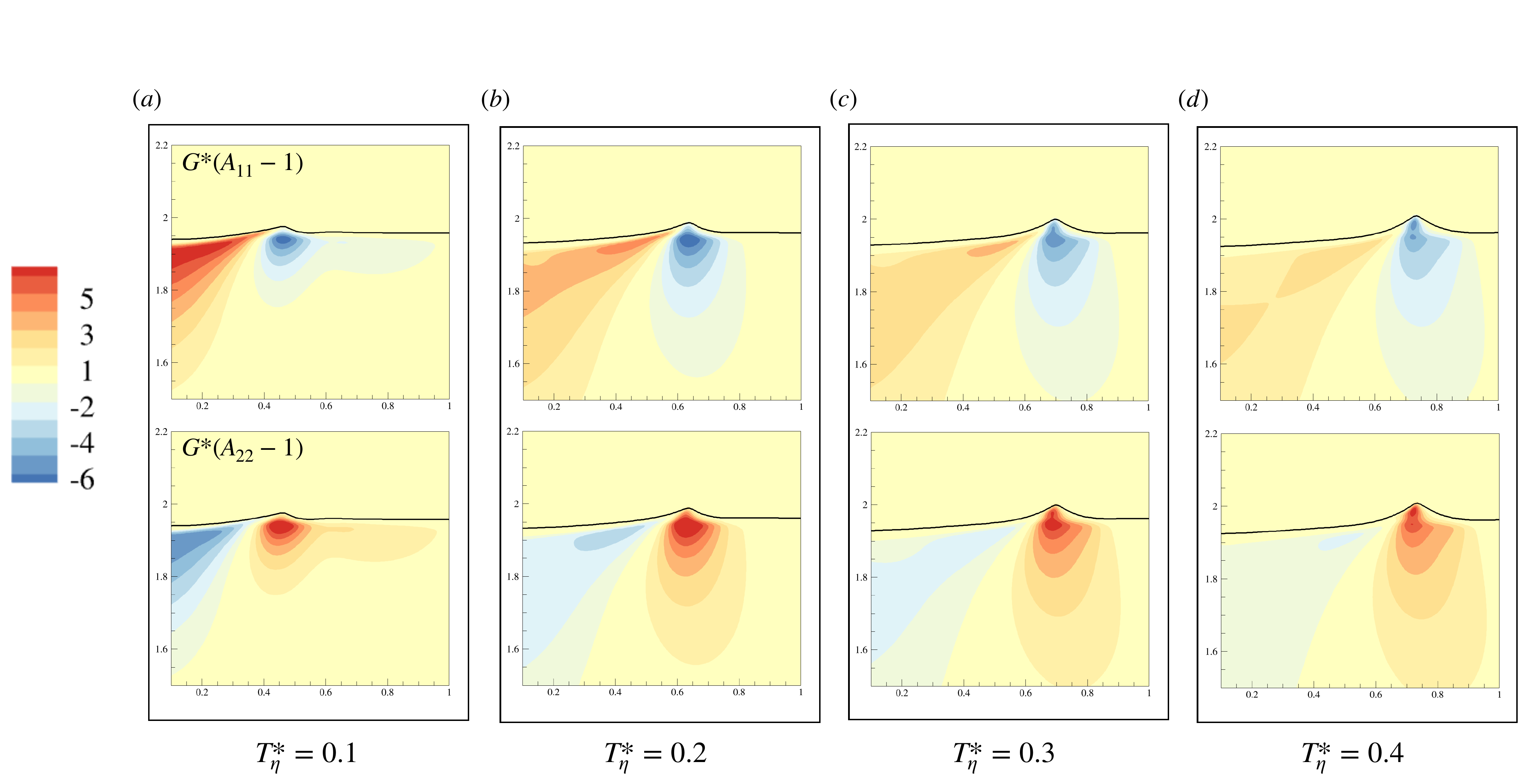}
    \caption{Scaled stress tensor component $\Pi_{11}^*$ (top), and $\Pi_{22}^*$ (bottom) of $\boldsymbol{\Pi}^*=G^*(\boldsymbol{A}-\boldsymbol{I})$ for $G^*=80$, $S_v^*=0.5$, $Oh=3.7$,  $T_\eta^*=[0.1,0.4]$. The strain tensor component $A_{11}-1$, and $A_{22}-1$ are in the range of [-0.075,0.063], and the deformation is small.\label{de10} 
    }
\end{figure*}

An advantage of the numerical simulation is that we are able to extract the evolution of the stress components. Further analysis of the contact line is based on the stress profile where Figures~\ref{de1000},~\ref{de10} provide the isocontour of the scaled stress tensor component $\Pi_{11}$ and $\Pi_{22}$, representing the two normal stress components. The values of the stress can be approximated from the left color map which ranges from [-0.16,0.36] for $G^*=0.8$, and [-6,5] for $G^*=80$. A large stress appears when the shear modulus is large. However, it is indicated that the conformation tensor $A$ is closer to $1$ when $G^*$ is large, which means the deformation at the contact line is small. This observation is consistent with figure~\ref{relax2}, that as the contact line moves, a self-similar wetting ridge can be observed for $G^*\gg1$.

%



\section{concluding remarks}\label{concluding}

In this study, we numerically investigated how viscoelasticity affects the interfacial flow when a Newtonian droplet comes in contact with an immicible viscoelastic film. As compared to the Newtonian case, a viscoelastic fluid has an influence on the motion of the droplet's center of mass. Numerical experiments establish the influence of the viscoelastic stress, achieved by manipulating polymer viscosity $\eta_p$ and polymer relaxation time $\lambda_p$. When elastic stresses are dominant, we show that the droplet dynamics become insensitive to the thickness of the viscoelastic film in both inertial and viscous flow regimes.

By focusing on the region around the contact line, we observed some intriguing behaviours linked to the elastic effects in the film. The spreading radius appears completely insensitive to changes in polymer relaxation time, but the vertical position of the contact line exhibits significant sensitivity to alterations in $G^*$. This is a consequence of the generated stresses, which in the radial growth of the droplet induces a shear flow, while the growth in height of the wetting ridge induces an extensional flow that promotes elastic effects. A noteworthy observation is the formation of sharper bridge profiles induced by polymers, particularly pronounced when the shear modulus $G^*\gg1$, consistent with the coalescence of two viscoelastic droplets~\cite{dekker2022elasticity}. Importantly, our simulations with different spreading factors reveal that while they may influence the final interface morphology, it does not significantly affect the evolution of the wetting ridge. Our findings help improve our understanding of the interplay between surface tensions, viscoelastic stress, and various influencing parameters in the spreading and engulfment of droplets by viscoelastic liquids.
\clearpage

\begin{acknowledgments}
We acknowledge the financial support of the Research Council of Norway through the program NANO2021 (project number 301138) and the PIRE project “Multi-scale, Multi-phase Phenomena in Complex Fluids for the Energy Industries”, founded by the Research Council of Norway and the National Science Foundation of USA under Award Number 1743794. This research was supported, in part, under National Science Foundation Grants: CNS-0958379, CNS-0855217, ACI-1126113, and OEC-2215760 (2022) and the City University of New York High Performance Computing Center at the College of Staten Island. The computations were also performed on resources provided by Sigma2 - the National Infrastructure for High-Performance Computing and Data Storage in Norway.
\end{acknowledgments}
\clearpage
\appendix

\section{Appendixes}

To effectively solve the conformation equation, there exist several methods i.e. finite difference method, the lattice Boltzmann method, and the logarithm method~\cite{alves2021numerical,wang2019lattice,hao2007simulation,comminal2015robust,lopez2019adaptive}. To validate our scheme and study the difference between different numerical methods, we show the comparison of the finite difference method and the logarithm method below. 

The discretized conformation equation of Eq.~\ref{gneq:5} can be shown as:
\begin{equation}
    \boldsymbol{A}^{n+1}=\boldsymbol{A}^n+\Delta t\left[\boldsymbol{A}(\nabla\boldsymbol{u}^{n})+
    (\nabla \boldsymbol{u}^{n})^T\boldsymbol{A} -
    \frac{1}{\lambda_p}\left(\boldsymbol{A}-\boldsymbol{I}\right)-\left(\boldsymbol{u}^{n}\cdot\nabla\right)\boldsymbol{A}\right].
\end{equation}
In order to construct a stable system, the Runge-Kutta method is employed, where $\boldsymbol{A}^{n+1}$ is approximated by an iterative method. We first evaluate four tensors of slopes  $\boldsymbol{K}_1$, $\boldsymbol{K}_2$, $\boldsymbol{K}_3$, $\boldsymbol{K}_4$ by:
\begin{equation}
    \boldsymbol{K}_1=-(\boldsymbol{u}^{n}\cdot\nabla)\boldsymbol{A}^n-
    \frac{1}{\lambda_p}\left(\boldsymbol{A}^n-\boldsymbol{I}\right)+\boldsymbol{A}^n(\nabla\boldsymbol{u}^{n})+
    (\nabla \boldsymbol{u}^{n})^T\boldsymbol{A}^n,
\end{equation}
\begin{align*}
    \boldsymbol{K}_2=&-(\boldsymbol{u}^{n}\cdot\nabla)(\boldsymbol{A}^n+0.5\boldsymbol{K}_1)-\\&
    \frac{1}{\lambda_p}\left((\boldsymbol{A}^n+0.5\boldsymbol{K}_1)-    \boldsymbol{I}\right)+(\boldsymbol{A}^n+0.5\boldsymbol{K}_1)(\nabla\boldsymbol{u}^{n})+
    (\nabla \boldsymbol{u}^{n})^T(\boldsymbol{A}^n+0.5\boldsymbol{K}_1),
\end{align*}
\begin{align*}
    \boldsymbol{K}_3=&-(\boldsymbol{u}^{n}\cdot\nabla)(\boldsymbol{A}^n+0.5\boldsymbol{K}_2)-\\&
    \frac{1}{\lambda_p}\left((\boldsymbol{A}^n+0.5\boldsymbol{K}_2)-    \boldsymbol{I}\right)+(\boldsymbol{A}^n+0.5\boldsymbol{K}_2)(\nabla\boldsymbol{u}^{n})+
    (\nabla \boldsymbol{u}^{n})^T(\boldsymbol{A}^n+0.5\boldsymbol{K}_2),
\end{align*}
\begin{align*}
    \boldsymbol{K}_4=&-(\boldsymbol{u}^{n}\cdot\nabla)(\boldsymbol{A}^n+\boldsymbol{K}_3)-\\&
    \frac{1}{\lambda_p}\left((\boldsymbol{A}^n+\boldsymbol{K}_3)-    \boldsymbol{I}\right)+(\boldsymbol{A}^n+\boldsymbol{K}_3)(\nabla\boldsymbol{u}^{n})+
    (\nabla \boldsymbol{u}^{n})^T(\boldsymbol{A}^n+\boldsymbol{K}_3).
\end{align*}
After the evaluation of those slopes, the $\boldsymbol{A}^{n+1}$ can be updated as:
\begin{equation}
    \boldsymbol{A}^{n+1}=\boldsymbol{A}^n+\frac{\Delta t}{6}\left(\boldsymbol{K}_1+2\boldsymbol{K}_2+2\boldsymbol{K}_3+\boldsymbol{K}_4\right).
\end{equation}
Typically, in the lattice Boltzmann method, the time step $\Delta t=1$. In addition to the numerical scheme for conformation tensor $\boldsymbol{A}$ we used in the manuscript, we tried the Logarithm scheme which is approved to be valid for high Weissenberg number problem~\cite{comminal2015robust,lopez2019adaptive}. The essential idea for the logarithm scheme is to alleviate the instability due to the advection-diffusion-reaction equation: 
\begin{equation}\label{log_1}
    \frac{\partial \boldsymbol{A}}{\partial t}+|\boldsymbol{u}|\frac{\partial\boldsymbol{A}}{\partial x}=\frac{1}{\lambda_p}\boldsymbol{A}.
\end{equation}
The stable condition of the discretized equation of Eq.~\ref{log_1} needs $\Delta x<\lambda_p|\boldsymbol{u}|$, where $\lambda_p$ has dimension time.
When the velocity magnitude $|\boldsymbol{u}|$ and the polymer relaxation time $\lambda_p$ are too small, high resolution is needed to resolve and stabilize the simulation. As we modify the equation to evolve the logarithm of conformation tensor $\boldsymbol{\psi}$ rather than direct evolve $\boldsymbol{A}$, where $\boldsymbol{\psi}=\log{\boldsymbol{A}}$, the above Eq.~\ref{log_1} becomes:
\begin{equation}\label{log_2}
    \frac{\partial \boldsymbol{\psi}}{\partial t}+|\boldsymbol{u}|\frac{\partial\boldsymbol{\psi}}{\partial x}=\frac{1}{\lambda_p}.
\end{equation}
In this case, the discretized equation of Eq.~\ref{log_2} is unconditionally stable. 
\begin{figure}[htb!]
  \centering
  \includegraphics[width=0.6\textwidth]{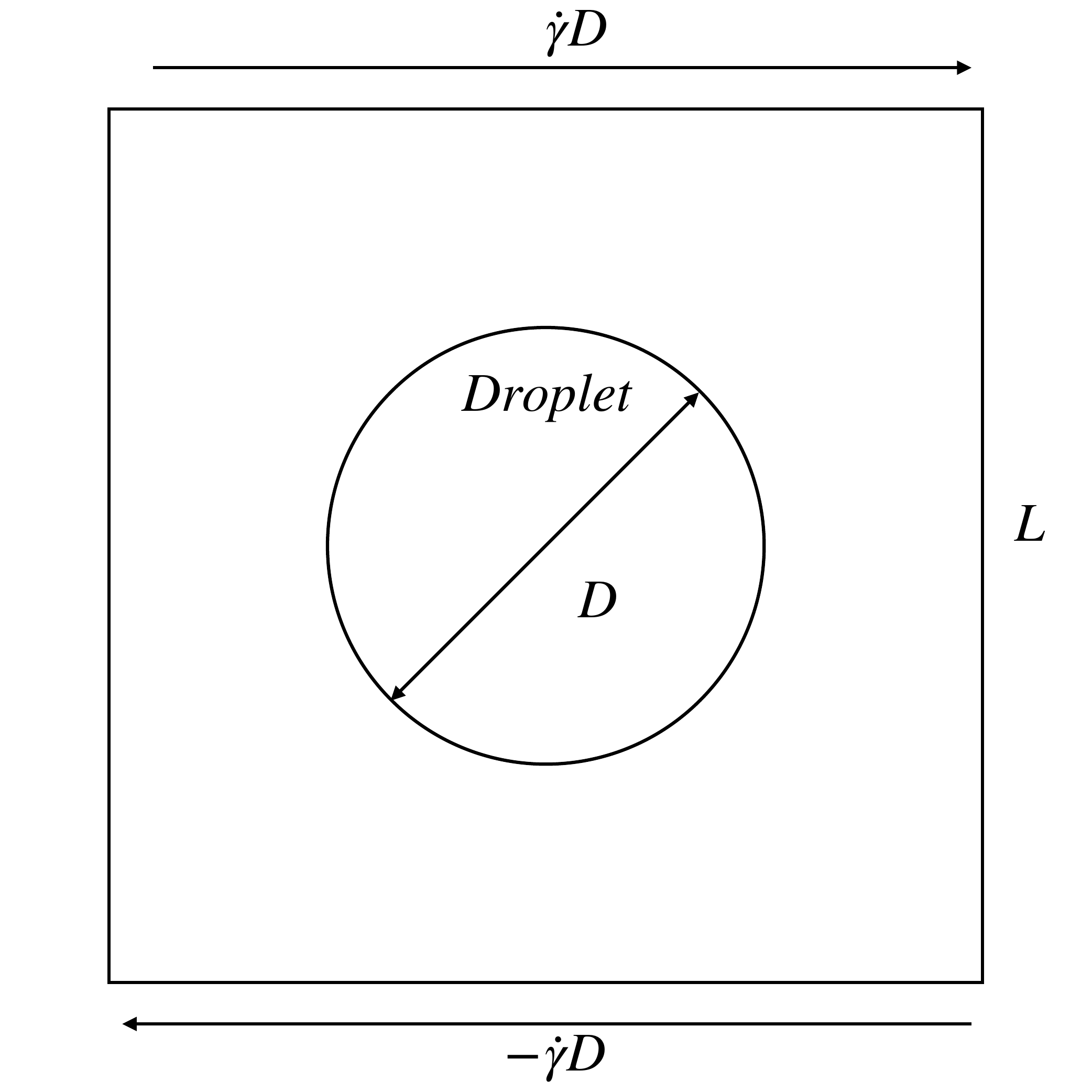}
    \caption{\label{shear_ini}Initial profile of the single droplet in a simple shear flow .}

\end{figure}
We here briefly introduce the numerical method, where a detailed derivation can be found in \cite{hao2007simulation}. Initially, the eigenvalue diagonal matrix $\boldsymbol{\Lambda}$ and eigenvector tensor $\boldsymbol{R}$ of the conformation tensor need to be evaluated, where:
\begin{equation}
    \boldsymbol{R}^T\boldsymbol{A}\boldsymbol{R}=\boldsymbol{\Lambda}.
\end{equation}
The transformation of strain tensor $\nabla \boldsymbol{u}$ then can be derived as:

\begin{equation}\label{grad_u}
    \nabla\boldsymbol{u}=\boldsymbol{\Omega}+\boldsymbol{B}+\boldsymbol{N}\boldsymbol{A}^{-1},
\end{equation}

 \begin{figure*}
\centering
  \includegraphics[width=\linewidth]{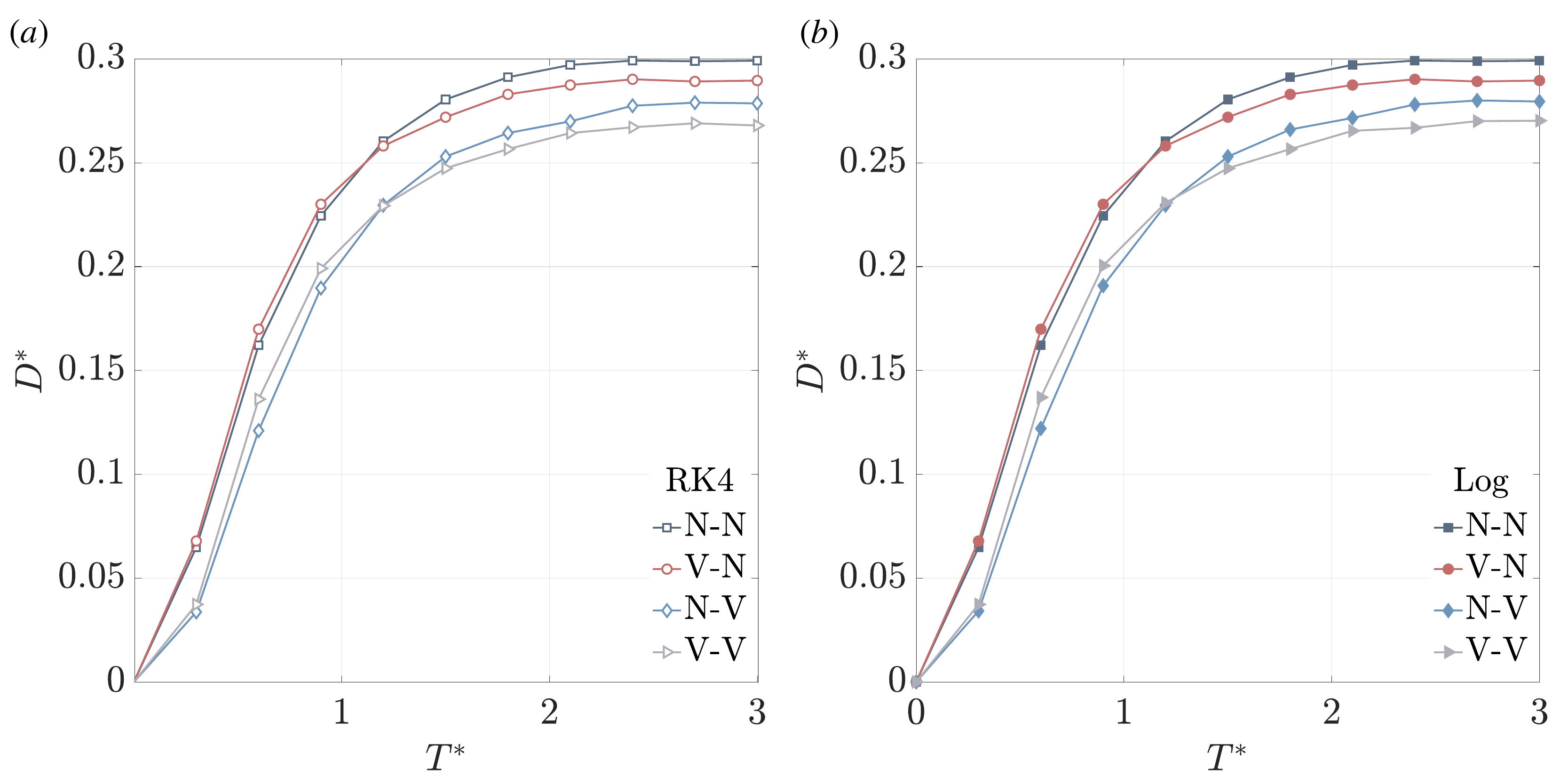}
    \caption{Evolution of the scaled length parameters in shear flow simulated by (a) finite difference scheme using RK4, and (b) Logarithm scheme. \label{shear} 
    }
\end{figure*}
and $(\nabla\boldsymbol{u})^T$ is also easily found. Tensor $\boldsymbol{\Omega}$, $\boldsymbol{B}$ and $\boldsymbol{N}$ can be computed by the tensor multiplication. Finally, after the transformation, we use the RK4 iterative scheme which we introduced in the section for the equation:
\begin{equation}
    \frac{\partial \boldsymbol{\psi}}{\partial t}+(\boldsymbol{u}\cdot\nabla)\boldsymbol{\psi}-(\boldsymbol{\Omega}\boldsymbol{\psi}-\boldsymbol{\psi}\boldsymbol{\Omega})-2\boldsymbol{B}=0.
\end{equation}
 \begin{figure*}
\centering
  \includegraphics[width=0.8\linewidth]{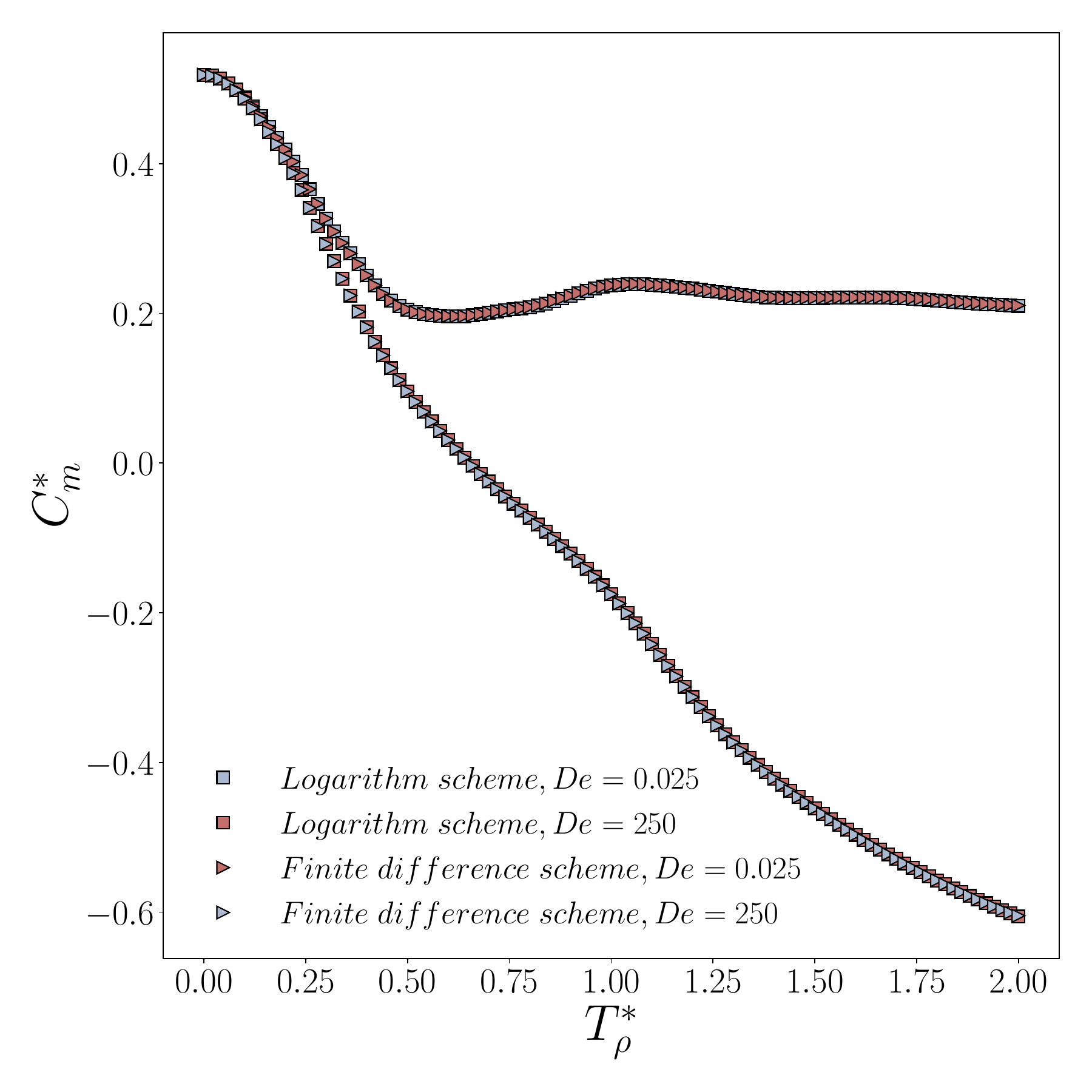}
    \caption{Comparsion between the logarithm scheme and the finite difference scheme. \label{test} 
    }
\end{figure*}

After we obtain $\boldsymbol{\psi}^{n+1}$, we transfer $\boldsymbol{\psi}$ back to $\boldsymbol{A}$, and introduce the reaction term $(\boldsymbol{A}-\boldsymbol{I})/\lambda_p$ to finalize the whole process. When the conformation tensor is computed, the same method is employed to introduce the viscoelastic stress to the momentum equation~\cite{zhao2023engulfment}.

We first validate the logarithm method and the finite difference method by a benchmark problem of the droplet deformation with a simple shear flow which is reported in \cite{wang2019lattice,chinyoka2005two}. The simulation setup is shown in Figure~\ref{shear_ini}. We place a droplet with diameter $D$, into a square with side length $L=2D$. In this case, the density ratio is set to $1$. A constant shear velocity $U=\dot{\gamma}D$ and $-U$ is applied to the top and bottom boundaries, and the left and right boundary conditions are set as periodic. We characteristic this problem by the Deborah number $De=\lambda_p\dot{\gamma}$, $Ca=\dot{\gamma}D\eta_m/\sigma$ and $Re=\rho\dot{\gamma}D^2/\eta_m$, where $\dot{\gamma}$ is the shear rate of the plates, and the $\eta_m$ is the viscosity of the matrix. There are four viscosity property definitions: (1) The matrix viscosity $\eta_m$, used to distinguish (2) the droplet viscosity $\eta_d$. (3) $\eta_p$ represents the viscosity of the polymer, and (4) $\eta_s$ denotes the viscosity of the solvent. In order to simplify this test, we only test for $\eta_p=\eta_s$. When we consider that the matrix is composed of viscoelastic fluid, $\eta_m=\eta_p+\eta_s=2\eta_d$. On the contrary, when the droplet is composed of viscoelastic fluid, $\eta_d=\eta_p+\eta_s=2\eta_m$ respectively.

Both the logarithm method and the finite difference method are applied to simulate four different cases: a Newtonian droplet in a Newtonian matrix (N-N), a Newtonian droplet in a viscoelastic matrix (N-V), a viscoelastic droplet in a Newtonian matrix (V-N), and a viscoelastic droplet in a viscoelastic matrix (V-V),  when $Ca=0.48$, $Re=1.2$, $De=0.4$. As shown in figure~\ref{shear}(a)(b), the temporal evolution of the length parameter $D^*=(a-b)/(a+b)$ when $T^*=\dot{\gamma}t=[0,3]$ for both numerical schemes are highly consistent and in good agreement with~\cite{wang2019lattice}, where $a$, and $b$ are the major and minor axes of the deformed droplet. The deformation of the droplet is relatively greater than that of the viscoelastic matrix when the matrix is made up of Newtonian fluid. Additionally, the deformation of the Newtonian droplet is greater than that of the viscoelastic droplet when the matrix fluid is the same.

Another test is performed to compare two numerical schemes for the three-phase problem. The same setup of the simulations as the section~\ref{engulf} is employed for both methods to evaluate the effect of $De$ for the interaction process. Figure.~\ref{test} shows the comparison of the mass center evolution with $De=[0.025, 250]$ for different numerical schemes. From the mass center evolution, the difference between the two schemes can be neglected.

\clearpage

\bibliography{main}

\end{document}